\documentclass[journal,draftcls,onecolumn,12pt,twoside]{IEEEtran}
\usepackage{graphicx}
\usepackage{subfigure}
\usepackage{amsfonts, amssymb, amsthm}
\usepackage{amsmath}
\usepackage{enumerate}
\usepackage{multirow}
\usepackage{color}
\usepackage{cite}
\usepackage{colortbl,booktabs}
\usepackage{diagbox}

\theoremstyle{definition} \newtheorem{theorem}{Theorem}
\theoremstyle{definition} \newtheorem{corollary}[theorem]{Corollary}
\theoremstyle{definition} \newtheorem{proposition}[theorem]{Proposition}
\theoremstyle{definition} 
\theoremstyle{definition} \newtheorem{lemma}[theorem]{Lemma}
\theoremstyle{definition} 

\theoremstyle{definition} 
\theoremstyle{definition} \newtheorem*{remark}{Remark}
\theoremstyle{definition} 
\theoremstyle{definition} 
{\end{list}}

\begin{document}
%

\title{Completion Delay of Random Linear Network Coding in Full-Duplex Relay Networks}
\author{\IEEEauthorblockN{Rina~Su\textsuperscript{\dag},~Qifu~Tyler~Sun\textsuperscript{\dag}{$^*$},~Zhongshan~Zhang\textsuperscript{\ddag}, ~Zongpeng~Li\textsuperscript{\S}} \\
\thanks{$^*~$ This paper was presented in part at 2021 IEEE International Symposium on Information Theory. R. Su and Q. T. Sun are with University of Science and Technology Beijing, Z. Zhang is with Beijing Institute of Technology, and Z. Li is with Tsinghua University. Q. T. Sun (Email:
qfsun@ustb.edu.cn) is the corresponding author.}
}

\maketitle
\thispagestyle{plain}
\pagestyle{plain}

\sloppy
\begin{abstract}
As the next-generation wireless networks thrive, full-duplex and relay techniques are combined to improve the network performance. Random linear network coding (RLNC) is another popular technique to enhance the efficiency and reliability of wireless communications. In this paper, in order to explore the potential of RLNC in full-duplex relay networks, we investigate two fundamental perfect RLNC schemes and theoretically analyze their completion delay performance. The first scheme is a straightforward application of conventional perfect RLNC studied in wireless broadcast, so it involves no additional process at the relay. Its performance serves as an upper bound for all perfect RLNC schemes. The other scheme allows sufficiently large buffer and unconstrained linear coding at the relay. It attains the optimal performance and serves as a lower bound for all RLNC schemes. For both schemes, closed-form formulae to characterize the expected completion delay at a single receiver as well as for the whole system are derived. Numerical results are also demonstrated to validate the theoretical characterizations, and compare the two fundamental schemes with the existing one. 
\end{abstract}
\section{Introduction}
\begin{figure}[t!]
\centering
\scalebox{0.36}
{\includegraphics[trim=10 13 10 5, clip]{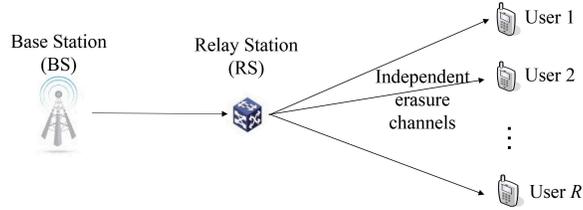}}
\caption{The system model of the relay network, in which the relay station is considered to be half-duplex in \cite{Chen2016}, and to be full-duplex in \cite{Chen&Meng2020} as well as in this paper.}
\label{fig:system_model}
\end{figure}

Among the emerging techniques for promoting evolution of the next generation (5G) wireless networks, relay techniques \cite{LiuGang2015} is widely considered for the purpose of catering to the ever-growing demand for throughput and coverage. As an up-and-coming paradigm, network coding (NC) and particularly random linear network coding (RLNC) \cite{Ho2006} has shown great capabilities to realize higher transmission efficiency and throughput in wireless communications. Different NC schemes have shown significant gains in a multitude of wireless transmission scenarios, from wireless broadcast \cite{Medard2008,Tsimbalo2018,Su&Sun2020, Ioannis2017, Atilla2013, Lucani2012} and wireless sensor networks \cite{Wang2012}, to D2D \cite{Huang2017} and Wi-Fi direct transmission \cite{Chieochan2011}. %
A considerable amount of research (e.g., \cite{Giacaglia2013,Pahlevani2014,Li2015,Pahlevani2020,Chen2016}) has investigated the combination of relay techniques with RLNC for the two-hop relay networks, which are the network model proposed in IEEE 802.16j and 3GPP LTE-Advanced \cite{Loa2010,Damnjanovic2011} for the sake of simplicity and explicitness of system design. %
In particular, in the two-hop relay network depicted in Fig. \ref{fig:system_model} with the relay station operating in the half-duplex mode, Ref. \cite{Chen2016} proposed an RLNC scheme with online scheduling that can achieve near-optimal completion delay performance, a key metric for transmission efficiency (e.g., \cite{Lucani2012},\cite{Giacaglia2013,Pahlevani2014,Li2015,Pahlevani2020,Chen2016},\cite{Sadeghi2010},\cite{Sameh2015}). %

Compared with the traditional half-duplex relay, the full-duplex relay can potentially double the throughput by simultaneously fetching and forwarding data \cite{Zhang2015,Zhang2016}. %
Moreover, with recent developments in the self-interference cancellation technology \cite{Y.Zhang2019}, the full-duplex relay has been considered as a key component to construct versatile networks in 5G. %
In the literature, there have been studies on combining full-duplex relaying with NC for performance enhancement over broadcast networks. For example, physical-layer NC has been applied in full-duplex relay networks with multiple receivers \cite{Han2018}. An NC-based Automatic Repeat Request (ARQ) scheme \cite{Zhu2019} was proposed to enhance the downlink throughput for a two-way full-duplex relay network. %
When the relay station in Fig. \ref{fig:system_model} operates in the full-duplex mode, the recent work \cite{Chen&Meng2020} proposed an RLNC scheme with scheduling, known as FBPF (Fewest Broadcast Packet First), which demonstrated a better throughput (equivalently, completion delay) performance than the ARQ. %
FBPF assumes full linear independence among the packets generated by the source, so it belongs to \emph{perfect} RLNC, which is always considered in analyzing the optimal performance RLNC can achieve (e.g., \cite{Medard2008,Shrader2011,Khamfroush2014,Giacaglia2013,Chen2016}). %
Even though FBPF is a type of perfect RLNC, it does not shed light on the best possible completion delay performance that RLNC can achieve because it does not fully utilize the packets buffered in the relay, \emph{i.e.}, it does not invoke coding while stores and forwards packets via an unnecessarily large buffer at the relay. %
In this paper, in order to further explore the potential of RLNC in the full-duplex relay network in Fig. \ref{fig:system_model}, we are inspired to investigate two fundamental perfect RLNC schemes and study their completion delay performance. The main contributions of this paper are summarized as follows.
\begin{itemize}
  \item We first investigate \emph{perfect RLNC without buffer}, which does not involve any buffer or additional process at the relay. As a result, this scheme provides a fundamental performance guarantee among all possible perfect RLNC schemes. For this scheme, explicit formulae of expected completion delay at a single receiver as well as for the system are deduced. %
  \item We then investigate \emph{perfect RLNC with buffer}, which allows sufficiently large buffer and unconstrained linear coding at the relay. It turns out that it attains the best completion delay performance among \emph{all} RLNC schemes. For this scheme, by modeling the transmission process as a Markov chain, we deduce a closed-form formula of the expected completion delay at a single receiver, which involves combinatorial numbers related to Schroeder paths. In order to compute the expected completion delay at a single receiver in a more handy manner, we further derive a recursive formula for it. %
  \item For perfect RLNC with buffer, we also model such a Markov chain that the expected system completion delay can be calculated by a formula built upon its $1$-step transition probability matrix, whose size grows exponentially with the increasing number of receivers. Furthermore, we characterize a non-trivial closed-form lower bound, which is the maximum of two individual ones that can be explicitly and recursively computed. The first stems from the expected system completion delay in wireless broadcast, and the other is selected to be the expected completion delay at a single receiver with the worst channel condition.
\end{itemize}

For full-duplex relay networks, the formulae of the expected completion delay of perfect RLNC without buffer obtained in this paper serve as upper bounds for the completion delay performance of \emph{all} perfect RLNC schemes, including FBPF \cite{Chen&Meng2020}, the relay-centered RLNC schemes \cite{Giacaglia2013}, and the one adapted from \cite{Chen2016}. %
The formulae of the expected completion delay of perfect RLNC with buffer obtained in this paper serves as lower bounds for the completion delay performance of \emph{all} RLNC schemes, including Fulcrum codes \cite{Fulcrum2018}, DSEP Fulcrum codes \cite{Fulcrum2020}, Sparse RLNC \cite{Tassi2016}, Telescopic codes\cite{Lucani2015}. The study of this paper provides a theoretical guideline for future works on the detailed design of RLNC-based transmission. 

We would like to remark that when the RS in the system operates in the half-duplex mode, the completion delay performance of perfect RLNC was analyzed in \cite{Chen2016}, but it does not shed light on the completion delay analyses in this paper, because its focus is to determine whether to fetch a packet from the BS or to broadcast a packet to the receivers at every timeslot at the RS.

The remainder of this paper is organized as follows. Section II introduces the system model and two fundamental perfect RLNC schemes. Section III theoretically analyzes the expected completion delay of two fundamental schemes. Section IV provides extensive numerical analyses to justify the theoretical characterizations and compare the two fundamental schemes' performance with FBPF. Section V concludes the paper.

Throughout the paper, we shall use $\mathbf{I}$ and $\mathbf{1}$ to respectively represent an identity matrix and an all-$1$ matrix, where the matrix size, if not explained, can be inferred in the context.
\section{System Model and Two Fundamental Perfect RLNC Schemes}
\subsection{System Model}
We consider the two-hop full-duplex relay network depicted in Fig. \ref{fig:system_model}, where a base station (BS) attempts to deliver $P$ packets to a set of $R$ receivers via a full-duplex relay station (RS) with a limited buffer size. The network transmission is considered to be time-slotted, that is, at every timeslot, the BS can deliver one packet to the RS, while the full-duplex RS can simultaneously fetch a coded packet from the BS and broadcast a coded packet to all receivers. The memoryless channel between the BS and the RS, together with the channel between the RS and every receiver $r$, are subject to independent random packet erasures with erasure probability $1-p_0$ and $1-p_r$, respectively. Every receiver is interested in recovering all $P$ original packets. Herein, the completion delay refers to the total number of packets transmitted by the BS before every receiver is able to recover all $P$ original packets. Notice that the definition of completion delay is the same as the one in \cite{Giacaglia2013}\cite{Chen&Meng2020}, which takes the initial $P$ packets transmitted by the BS into account, so it is slightly different from the one in \cite{Giacaglia2013}\cite{Su&Sun2020}. The packet number $P$ divided by the completion delay is set as a measurement of throughput in \cite{Chen&Meng2020}.

In practice, at every timeslot, even though the BS and the RS transmit simultaneously, the RS obtains what the BS transmits at the end of the timeslot. Thus, for the RS, the broadcast process is always one timeslot behind the reception process, \emph{i.e.}, what the RS broadcasts at timeslot $j$ has nothing to do with what it receives at timeslot $j$. Same as the consideration in  \cite{Giacaglia2013}, in this paper, we shall also ignore this constant timeslot lag at the RS and this assumption does not affect the analysis on the completion delay performance.

For the full-duplex relay network, Ref. \cite{Chen&Meng2020} has proposed the FBPF scheduling scheme, in which if the RS buffer is not empty, the RS selects a packet that has been broadcast the fewest number of times from the unlimited buffer, and broadcasts the selected packets to the receivers. Since perfect RLNC is adopted in the design of the FBPF scheme, any $P$ different packets received by a receiver are assumed to be linearly independent. %
\subsection{Two Fundamental Perfect RLNC Schemes}
Since the FBPF perfect RLNC scheme permits unlimited buffer at the RS for additional scheduling procedures, it is not a perfect RLNC scheme with the simplest setting at the RS. Thus, it does not reflect a fundamental performance guarantee provided by perfect RLNC in the full-duplex relay network, that is, there exist other perfect RLNC schemes with simpler settings at the RS and their completion delay performance is not as good as that of the FBPF scheme. On the other hand, the FBPF perfect RLNC scheme cannot yield the best performance gain in terms of completion delay as it does not involve NC at the RS.%

In order to study the fundamental completion delay performance of perfect RLNC in the full-duplex relay network, we consider two basic perfect RLNC schemes, one without buffering at the RS and the other with buffering and recoding at the RS.

The first scheme, called perfect RLNC without buffer, does not require buffer and thus there is no recoding at the RS. The role of the RS is just to directly forward the received packets. At every timeslot, the BS delivers one coded packet, which is a random linear combination of $P$ original packets, to the RS. If the RS successfully receives the coded packet, then it will broadcast it to all receivers. Otherwise, the RS will not transmit anything. %
Notice that the full-duplex RS can simultaneously fetch the coded packet from the BS and broadcast it to all receivers. As perfect RLNC is considered, any $P$ coded packets generated by the BS are assumed linearly independent and sufficient to recover the $P$ original packets. As a result, once every receiver obtains $P$ packets, the transmission completes. Since this scheme does not involve any extra operation at the RS, it is the most straightforward and simplest application of perfect RLNC in the full-duplex relay network. As a result, it provides a fundamental performance guarantee for all perfect RLNC schemes designed for the full-duplex relay network in terms of completion delay.%

The other scheme, called perfect RLNC with buffer, assumes buffer size $P$ and no coding constraints at the RS. At each timeslot, the BS transmits a coded packet to the RS, and if the RS receives the packet and its buffer is not full, then it stores the received packet in its buffer. Meanwhile, no matter whether the RS successfully receives the packet from the BS or not, it  broadcasts a packet which is a random linear combination of all the packets stored in the buffer. %
Due to the causality at the RS to firstly buffer received packets and then broadcast a linear combination of buffered packets to receivers, the number of linearly independent packets obtained at a receiver is always no larger than the number of packets buffered at the RS. Once every receiver obtains $P$ linearly independent packets, the transmission completes. %
As perfect RLNC is considered, the $P$ original packets can be recovered from any $P$ coded packets generated by the BS. Moreover, there is no coding constraint at the RS in this scheme. As a result, as long as the RS stores $P$ coded packets transmitted from the BS, the $P$ original packets can be recovered from any $P$ randomly generated coded packets at the RS. This justifies why setting the buffer size to $P$ is sufficient, and actually, when the RS successfully stores $P$ coded packets transmitted from the BS in the buffer, there is no need for the BS to transmit new coded packets to the RS at all. To sum up, the scheme perfect RLNC with buffer attains the best completion delay performance among all perfect RLNC schemes in the full-duplex relay network.

\section{Completion Delay Analysis}
One of the main contributions in this paper is to theoretically analyze the expected completion delay of the benchmark schemes introduced in Sec. II-B, that is, the perfect RLNC scheme without buffer and with buffer, respectively.

\subsection{Perfect RLNC without Buffer}
In perfect RLNC without buffer, the completion delay at a single receiver $r$, denoted by $D_{0, r}$, is defined to be the number of packets the BS transmits till receiver $r$ is able to recover all $P$ original packets, and the completion delay for the system, denoted by $D_{0}$, is defined as
\[
D_0 = \max\{D_{0,1}, D_{0,2}, \ldots, D_{0,R}\}.
\]

The completion delay $D_{0,r}$ at a single receiver $r$ follows the negative binomial distribution with the probability mass function $\mathrm{Pr}(D_{0,r} = P + d) = \binom{P+d-1}{P-1} (p_0p_r)^P(1-p_0p_r)^d, d \geq 0$, so that the expectation of $D_{0,r}$ is equal to
\begin{equation}
\label{eq:delay without_buffer_single_receiver}
\mathbb{E}[D_{0,r}] = P/(p_0p_r).
\end{equation}

\begin{proposition}
\label{thm:perfect scheme without_buffer}
The expected completion delay of the perfect RLNC scheme without buffer is
\begin{equation}
\label{eqn:perfect scheme without_buffer}
\mathbb{E}[D_0] = \frac{1}{p_0}(P+\sum\nolimits_{d \geq 0} (1 - \prod\nolimits_{1 \leq r \leq R} I_{p_r}(P, d+1) )),
\end{equation}
where $I_{p_r}(P,d+1) = \sum_{j=0}^{d} \binom{P+j-1}{P-1} p_{r}^P(1-p_r)^{j}$ is the regularized incomplete beta function.
\end{proposition}
\begin{IEEEproof}
Let $\hat{D}_P$ denote the number of packets broadcast from the RS till every receiver is able to decode all $P$ original packets. Since the considered perfect RLNC scheme does not have any buffer at the RS, any $P$ out of the $\hat{D}_P$ packets broadcast from the RS are linearly independent. Thus, $\hat{D}_P$ can be regarded as the completion delay for the wireless broadcast system with $P$ original packets, $R$ receivers and independent erasure probability $1 - p_r$. By Theorem 1 in \cite{Su&Sun2020},
\begin{equation}
\label{eqn:perfect_for_wireless_broadcast}
\mathbb{E}[\hat{D}_P] = P+\sum\nolimits_{d \geq 0} (1 - \prod\nolimits_{1 \leq r \leq R} I_{p_r}(P, d+1))
\end{equation}
On the other hand, as the RS immediately broadcasts a packet it successfully receives from the BS, $\hat{D}_P$ also represents the number of successfully received packets from the BS at the RS, and $D_0$ just represents the number of transmissions from the BS till the RS successfully receives $\hat{D}_P$ packets. As it takes on average $1/p_0$ transmissions to successfully receive one packet at the RS from the BS, $\mathbb{E}[D_0]=\mathbb{E}[\hat{D}_P]/p_0$, which implies \eqref{eqn:perfect scheme without_buffer} based on \eqref{eqn:perfect_for_wireless_broadcast}.
\end{IEEEproof}

For the special case that the channel from the RS to every receiver $r$ is perfect, \emph{i.e.}, $p_r = 1$, the full-duplex relay network becomes essentially the same as a point-to-point transmission, so the system completion delay of both perfect RLNC schemes considered herein follows the negative binomial distribution with the expected value $\frac{P}{p_0}$. %
For the other special case $p_0 = 1$, the full-duplex relay network degenerates to the wireless broadcast with erasure probability $1-p_r$, so $\mathbb{E}[D_0]$ is given by $P+\sum\nolimits_{d \geq 0} (1 - \prod\nolimits_{1 \leq r \leq R} I_{p_r}(P, d+1))$, same as the one obtained in \cite{Su&Sun2020}.

\subsection{Perfect RLNC with Buffer, Single Receiver Case}
In perfect RLNC with buffer, the completion delay at receiver $r$ is denoted by $D_{P, r}$, where $P$ means the number of original packets generated by the source and to be recovered at every receiver. The system completion delay, denoted by $D_{P}$, is defined as
\[
D_P = \max\{D_{P,1}, D_{P,2}, \ldots, D_{P,R}\}.
\]

In order to characterize the expected completion delay $\mathbb{E}[D_{P, r}]$, we first recall the following combinatorial number
\begin{equation}
T_{i,j} = \frac{1}{j+1}\binom{i+j}{i}\binom{i}{j},\quad 0 \leq j \leq i
\end{equation}
The Schroeder path (see, e.g., \cite{Schroeder_path}) from $(0,0)$ to $(i,i)$ is a path with possible movement $(+1, 0)$, $(0, +1)$, $(+1, +1)$ in every step transition and with $x \geq y$ for every point $(x,y)$ in the path. Then, $T_{i,j}$ represents the number of Schroeder paths from $(0,0)$ to $(i, i)$ with $i+j$ step transitions.

\begin{theorem}
\label{thm:perfect scheme with_buffer}
At a single receiver $r$, the expected completion delay of the perfect RLNC scheme with buffer is
\begin{equation}
\label{eqn:thm_w_buffer}
\mathbb{E}[D_{P,r}] = \frac{P}{p_0}+\frac{P}{p_r}- 1 + \sum_{i=0}^{P-2}\sum_{j=0}^{i} \frac{(P-i-1)T_{i,j}(p_0p_r)^i}{(p_0p_r - p_0 - p_r)^{i+j+1}}.
\end{equation}
\begin{IEEEproof}
Model the transmission process as a Markov chain $\mathcal{M}_P$ consisting of $\frac{(P+1)(P+2)}{2}$ states. Every state, labeled as $(i, j)$, represents the scenario that the RS and the receiver have respectively obtained $i$ and $j$ packets. Due to the causality at the RS to firstly buffer received packets (from the BS) and then broadcast a random linear combination of the buffered packets to receivers, all states $(i,j)$ in $\mathcal{M}_P$ have $0 \leq j \leq i \leq P$, and the only absorbing state in $\mathcal{M}_P$ is $(P,P)$ (assuming $p_0, p_1, \ldots, p_R \neq 0$). %
The $1$-step transition probability $p_{ij,i'j'}$ from a transient state $(i,j)$ to another state $(i', j')$ is given by
\begin{itemize}
\label{eqn:transition_probability_perfect_RLNC_with_buffer}
\item for $0 \leq j = i < P$, $p_{ij,ij} = 1-p_0, p_{ij, (i+1)j} = p_0(1-p_r), p_{ij,(i+1)(j+1)} = p_0p_r$;
\item for $0 \leq j < i < P$, $p_{ij,ij}=(1-p_0)(1-p_r)$,$p_{ij,(i+1)j}=p_0(1-p_r)$,$p_{ij,i(j+1)}=(1-p_0)p_r$,$p_{ij,(i+1)(j+1)}=p_0p_r$.
\item for $0 \leq j < i = P$, $p_{ij,ij}=1-p_r$,$p_{ij,i(j+1)}=p_r$.
\end{itemize}%

Let $\mathbf{P}$ denote the matrix of $1$-step transition probabilities among all $\frac{(P+1)(P+2)}{2}-1$ transient states. Assume the states are ordered lexicographically, so that the first row/column in $\mathbf{P}$ is indexed by the state $(0, 0, \ldots, 0)$. By the standard technique to calculate the expected transition times from a transient state to an absorbing one (see, e.g., Sec. 4.6 in \cite{SheldonRoss}), the expected completion delay $\mathbb{E}[D_{P,r}]$ can be formulated as
\begin{equation}
\label{eqn:expectation_w_buffer}
\mathbb{E}[D_{P,r}] = (1,0,\ldots,0)(\mathbf{I}-\mathbf{P})^{-1}\mathbf{1},
\end{equation}
where $(1,0,\ldots,0)$ represents the $\left(\frac{(P+1)(P+2)}{2}-1\right)$-dimensional row unit vector with the first entry equal to $1$. The details to derive \eqref{eqn:thm_w_buffer} based on \eqref{eqn:expectation_w_buffer} is provided in Appendix-\ref{proof_of eq41}.
\end{IEEEproof}
\end{theorem}

\begin{remark}
In the literature, Ref. \cite{Lucani2012, Giacaglia2013, Medard2011, Fulcrum2018,Nistor2011} also studied the completion delay performance from a Markov chain approach in different network settings. For example, \cite{Lucani2012} characterized the system completion delay for wireless broadcast with feedback by means of a moment generating function. For wireless broadcast with $2$ receivers, Ref. \cite{Nistor2011} characterized the cumulative distribution function of the system completion delay of RLNC by means of a Markov chain model. For the relay-centered network with a single receiver, Ref. \cite{Giacaglia2013} established a Markov chain model to obtain an implicit recursive formula for the expected completion delay of perfect RLNC. %
However, in the above references, there is \emph{no explicit} expression for the completion delay studied. In comparison, we also adopt the Markov chain approach to model the transmission process in full-duplex relay networks, and by deliberate analysis we obtain a closed-form formula of the completion delay at a single receiver. \hfill $\blacksquare$
\end{remark}

With $P$ increasing, the calculation of $\mathbb{E}[D_{P,r}]$ according to (\ref{eqn:thm_w_buffer}) becomes tedious because it involves the combinatorial number $T_{i,j}$, which can be extremely large. Stemming from (\ref{eqn:thm_w_buffer}), we next deduce an equivalent recurrence expression for $\mathbb{E}[D_{P,r}]$.

For $i \geq 0$, define
\begin{equation}
\label{eqn:Bi_definition}
B(i) =  \sum\nolimits_{j = 0}^{i} \frac{T_{i,j}(p_0p_r)^i}{(p_0p_r - p_0 - p_r)^{i+j+1}}.
\end{equation}
Thus, (\ref{eqn:thm_w_buffer}) can be rewritten as
\begin{equation}
\label{eqn:E_D_P_r}
\mathbb{E}[D_{P,r}] = \frac{P}{p_0}+\frac{P}{p_r}-1+\sum_{i=0}^{P-2}(P-i-1)B(i).
\end{equation}
Moreover, one can readily see
\begin{equation}
\label{eqn:P_P_1}
\mathbb{E}[D_{P+1,r}]-\mathbb{E}[D_{P,r}] = \frac{1}{p_0}+\frac{1}{p_r}+\sum_{i=0}^{P-1}B(i).
\end{equation}
\begin{corollary}
$B(i)$ can be recursively expressed as
\begin{equation}
\label{eqn:B_i}
B(i) = -\frac{p_0p_r}{\Delta}\left(B(i-1) + \sum_{j=0}^{i-1}B(j)B(i-j-1)\right).
\end{equation}
with the initial value $B(0) = -\frac{1}{p_0 + p_r - p_0p_r}$.
\begin{IEEEproof}
Same as in the proof of Theorem \ref{thm:perfect scheme with_buffer}, write $\Delta = p_0 + p_r - p_0p_r = 1 - (1-p_0)(1-p_1)$ for short. Thus, $B(i)$ can be expressed as
\begin{equation}
\label{eqn:B_i2}
B(i) = -\frac{1}{\Delta} \sum\nolimits_{j = 0}^{i} T_{i,j} \left(-\frac{p_0p_r}{\Delta}\right)^{i-j}\left(\frac{p_0p_r}{\Delta^2}\right)^j.
\end{equation}
First, it is trivial to see $B(0) = -1/\Delta$.

Recall that $T_{i,j}$ represents the number of Schroeder paths from $(0,0)$ to $(i,i)$ with exactly $i+j$ step transitions. Moreover, for every Schroeder path in $T_{i,j}$, the number of step transitions that are in the form of from $(i',j')$ to $(i'+1,j'+1)$, from $(i',j')$ to $(i'+1,j')$ and from $(i',j')$ to $(i',j'+1)$ are respectively equal to $i-j$, $j$ and $j$. Now assign a weight to every step transition as follows. If the step transition is from $(i',j')$ to $(i'+1,j'+1)$, then its weight is $-\frac{p_0p_r}{\Delta}$; if the step transition is from $(i',j')$ to $(i'+1,j')$ or from $(i',j')$ to $(i',j'+1)$, then its weight is $\frac{\sqrt{p_0p_r}}{\Delta}$. %
For every Schroeder path, define its weight as the \emph{product} of all the weights for the step transitions in the path. Thus, based on \eqref{eqn:B_i2}, $-\Delta B(i)$ can be regarded as the sum of weights of all Schroeder paths from $(0,0)$ to $(i,i)$.

All Schroeder paths from $(0,0)$ to $(i,i)$ can be partitioned into $i+1$ categories. The first category consists of all those Schroeder paths that contain the step transition from $(0, 0)$ to $(1, 1)$, which has weight $-\frac{p_0p_r}{\Delta}$. Thus, the sum of weights of all Schroeder paths in the first category is equal to $-\frac{p_0p_r}{\Delta}(-\Delta B(i-1))= p_0p_rB(i-1)$. For $0 \leq j \leq i-1$, the $(j+1)^{st}$ category consists of all those Schroeder paths that satisfy the followings:
\begin{itemize}
\item the step transitions from $(0, 0)$ to $(1, 0)$ and from $(j+1,j)$ to $(j+1,j+1)$ are contained;
\item the step transitions from $(1,0)$ to $(j+1, j)$ is equivalent to a Schroeder path from $(0,0)$ to $(j,j)$;
\item the step transitions from $(j+1,j+1)$ to $(i,i)$ is equivalent to a Schroeder path from $(0,0)$ to $(i-j-1,i-j-1)$.
\end{itemize}
As a result, the sum of weights of all Schroeder paths in category ${j+1}$ equals to $\frac{p_0p_r}{\Delta^2}(-\Delta B(j))(-\Delta B(i-j-1)) = p_0p_rB(j)B(i-j-1)$. To add up the weights of all Schroeder paths in all $i+1$ categories, we obtain
\[
-\Delta B(i) = p_0p_rB(i-1) + \sum\nolimits_{j=0}^{i-1}p_0p_rB(j)B(i-j-1),
\]
that is, \eqref{eqn:B_i} holds.
\end{IEEEproof}
\end{corollary}
Based on \eqref{eqn:P_P_1} and \eqref{eqn:B_i}, $\mathbb{E}[D_{P+1,r}]$ can be recursively computed for arbitrarily large $P$. %
It is also interesting to observe from Eq. \eqref{eq:delay without_buffer_single_receiver}, \eqref{eqn:thm_w_buffer}, \eqref{eqn:P_P_1} and \eqref{eqn:B_i} that the parameters $p_0$ and $p_r$ have the same impact on the expected completion delay at a single receiver $r$ for both perfect RLNC without buffer and perfect RLNC with buffer.

\subsection{Perfect RLNC with Buffer, Multiple-Receiver Case}
\begin{figure}[t!]
\centering
\scalebox{0.36}
{\includegraphics[trim=10 10 10 10, clip]{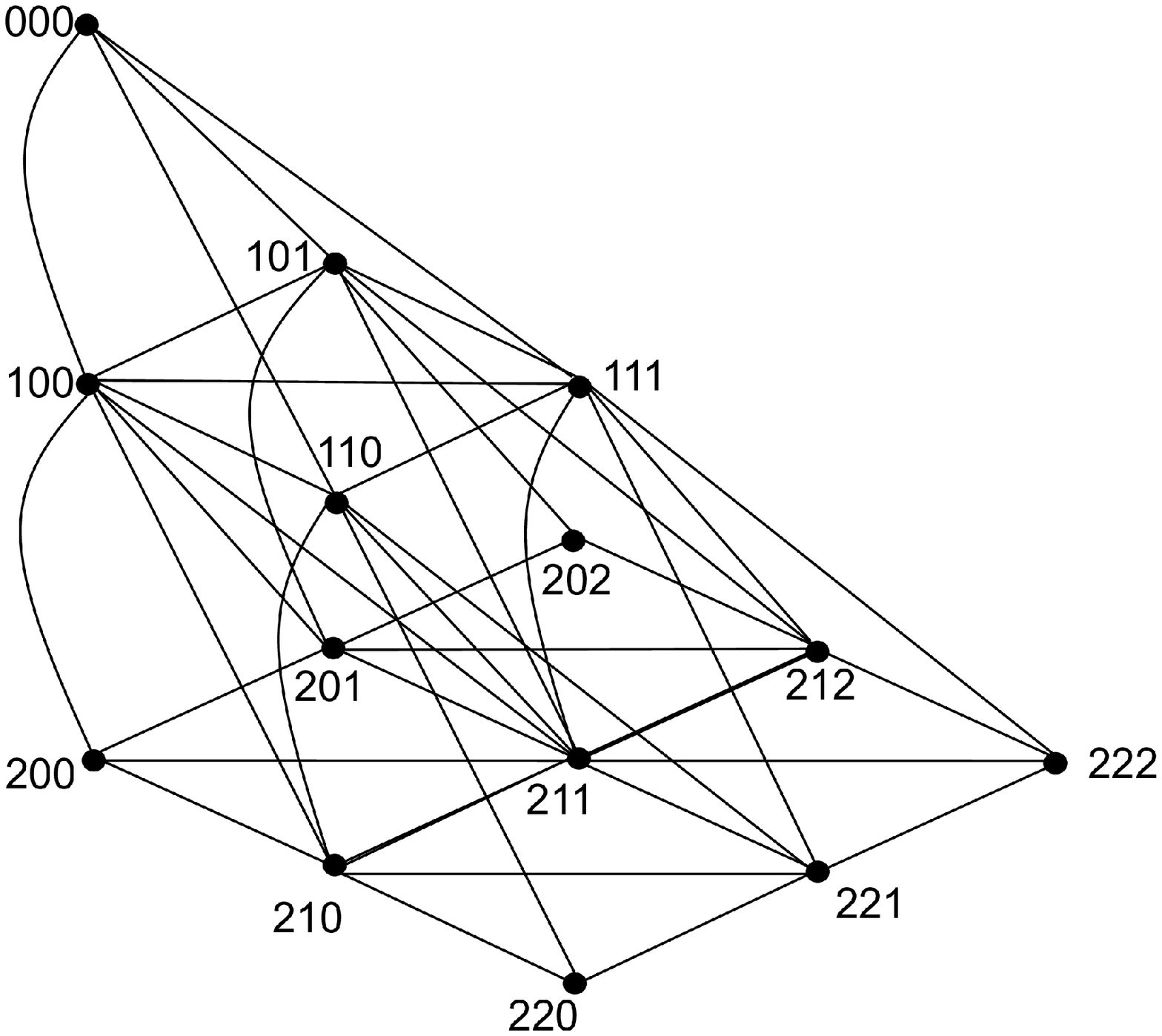}}
\caption{An illustration of the $1$-step transition diagram for the Markov chain $\mathcal{M}_{P,R}$ with $R = P = 2$. There are $1^2+2^2+3^2 = 14$ states, all states except for $(2,2,2)$ are transient, and there is a state transition from every transient state to itself. For brevity, every state $(s_0, s_1,s_2)$ is labeled as $s_0s_1s_2$, the transition from every transient state to itself is not depicted, and the edge direction is not marked. Specifically, whenever there is an edge between $\mathbf{s} = (s_0, s_1,s_2)$ and $\mathbf{s}' = (s_0',s_1',s_2')$ with $s_j \leq s_j'$ for all $0 \leq j \leq 2$, it represents a transition from $\mathbf{s}$ to $\mathbf{s}'$.}
\label{fig:markov_chain}
\end{figure}

The transmission process of the perfect RLNC scheme with buffer for multiple receivers on the full-duplex relay network can be modeled as a Markov chain, denoted by $\mathcal{M}_{P,R}$, in the following way. Every state in $\mathcal{M}_{P,R}$ can be labeled by an $(R+1)$-tuple $\mathbf{s} = (s_0, s_1,s_2,\ldots,s_R)$, where $s_0$ represents the number of packets successfully received by the RS, and $s_r$ represents the number of packets successfully received by receiver $r$. Notice that $0\leq s_r \leq s_0 \leq P$ for every receiver $r$. Thus, by conditioning on $s_0$, we can compute the number of states in the Markov chain $\mathcal{M}_{P,R}$ as $\sum_{s_0 = 0}^{P} (s_0+1)^R$ states. %
Except for the state $(P,P,\ldots,P)$, which is absorbing, all other $\sum_{s_0 = 0}^{P} (s_0+1)^R -1$ states are transient (assuming $p_0, p_1, \ldots, p_R \neq 0$). There is a $1$-step transition in the Markov chain once the BS broadcasts a new packet in a timeslot. %
An illustration of the $1$-step transition diagram for the case $R = P = 2$ is given in Fig. \ref{fig:markov_chain}. %
We next define the $1$-step transition probability from state $\mathbf{s} = (s_0, s_1,s_2,\ldots,s_R)$ to state $\mathbf{s}' = (s_0',s_1',s_2',\ldots,s_R')$ for the Markov chain. %

Let $\mathcal{R}$ denote the set of receivers who have not obtained $s_0$ packets at state $\mathbf{s}$ yet, that is, $\mathcal{R} = \{1 \leq r \leq R: s_r < s_0\}$. In addition, denote by $\mathcal{R}'$ the set of receivers who have obtained a new packet after the $1$-step transition from $\mathbf{s}$ to $\mathbf{s}'$, that is, $\mathcal{R}' = \{r \in \mathcal{R}: s_r' = s_r + 1\}$. %
The $1$-step transition probability from $\mathbf{s}$ to $\mathbf{s}'$ can be formulated by the following $3$ different cases depending on the value of $s_0$ and $s_0'$.
\begin{itemize}
\item Case $1$: $s_0' = s_0 < P$. In this case $\mathcal{R}' \subseteq \mathcal{R}$. We have
    \begin{equation}
    \label{eq:transition_probability_case_1}
    p_{\mathbf{s},\mathbf{s}'} = (1-p_0)\left(\prod\nolimits_{r\in \mathcal{R'}}p_r\right) \left(\prod\nolimits_{r\in \mathcal{R}\backslash\mathcal{R'}}(1-p_r)\right).
\end{equation}

\item Case $2$: $s_0' = s_0 = P$. In this case $\mathcal{R}' \subseteq \mathcal{R}$. We have
    \begin{equation}
    \label{eq:transition_probability_case_2}
    p_{\mathbf{s},\mathbf{s}'} = \left(\prod\nolimits_{r\in \mathcal{R'}}p_r\right) \left(\prod\nolimits_{r\in \mathcal{R}\backslash\mathcal{R'}}(1-p_r)\right).
\end{equation}

\item Case $3$: $s_0' = s_0 + 1$. Notice that $\mathcal{R}'$ is not necessarily contained in $\mathcal{R}$ in this case. We have
\begin{equation}
    \label{eq:transition_probability_case_3}
    p_{\mathbf{s},\mathbf{s}'} = p_0\left(\prod\nolimits_{r\in \mathcal{R'}}p_r\right) \left(\prod\nolimits_{r\notin \mathcal{R'}}(1-p_r)\right).
\end{equation}
\end{itemize}
Analogous to \eqref{eqn:expectation_w_buffer}, the expected completion delay for the system can be expressed as
\begin{equation}
\label{eqn:E_D_P}
\mathbb{E}[D_P] = (1,0,\ldots,0)(\mathbf{I}-\mathbf{P})^{-1}\mathbf{1},
\end{equation}
where $\mathbf{P}$ denotes the matrix of $1$-step transition probabilities among all transient states, with the first row/column indexed by the state $(0, \ldots, 0)$. %

Since the number of states in the Markov chain $\mathcal{M}_{P,R}$ increases exponentially with increasing $R$, it may not be convenient to compute $\mathbb{E}[D_P]$ based on \eqref{eqn:E_D_P}. %
We next provide an alternative way to analyze $\mathbb{E}[D_P]$ by exploring its connection with $\mathbb{E}[\hat{D}_{P}]$,
which is given by \eqref{eqn:perfect_for_wireless_broadcast} and denotes the expected system completion delay for the special case $p_0 = 1$, or equivalently, the expected system completion delay in wireless broadcast. For this purpose, we first focus on a single receiver $r$ and compare $\mathbb{E}[D_{P,r}]$ with $\mathbb{E}[\hat{D}_{P,r}] = P/p_r$, which represents the expected completion delay at receiver $r$ for the special case $p_0 = 1$. %
For $1 \leq j \leq P$, let $S_j$ and $T_{j,r}$ respectively denote the number of timeslots that the RS and receiver $r$ take to receive the $j^{th}$ packet. Thus, $S_1 < \ldots < S_P$, $T_{1,r} < \ldots < T_{P,r}$, and $S_j \leq T_{j,r}$ for all $1 \leq j \leq P$. By conditioning on the relation between $S_{P+1}$ and $T_{P, r}$, we can deduce the followings.
\begin{itemize}
\item $S_{P+1} \leq T_{P,r}$, which means the RS has received the $(P+1)^{st}$ packet upon the reception of the $P^{th}$ packet by receiver $r$. In this case, it takes on average $\frac{1}{p_r}$ timeslots for receiver $r$ to further obtain the $(P+1)^{st}$ packet so that $\mathbb{E}[D_{P+1,r}] - \mathbb{E}[D_{P,r}] = \frac{1}{p_r} = \mathbb{E}[\hat{D}_{P+1,r}] - \mathbb{E}[\hat{D}_{P,r}]$;
\item $S_{P+1} > T_{P,r}$, which means upon the reception of the $(P+1)^{st}$ packet at the RS, receiver $r$ has only received fewer than $P$ packets. In this case, it takes on average $\frac{1}{p_0}$ additional timeslots for the RS to receive the $(P+1)^{st}$ packet, and $\frac{1}{p_r}-1$ timeslots for receiver $r$ to receive the $(P+1)^{st}$ packet. Thus, $\mathbb{E}[D_{P+1,r}] - \mathbb{E}[D_{P,r}] = \frac{1}{p_0}+\frac{1}{p_r}-1 = \mathbb{E}[\hat{D}_{P+1,r}] - \mathbb{E}[\hat{D}_{P,r}] + \frac{1}{p_0} - 1$.
\end{itemize}
In all, we have
\begin{equation}
\label{eqn:D_Pr_D_Pr_hat_relation1}
\mathbb{E}[D_{P+1,r}] - \mathbb{E}[D_{P,r}] = \mathbb{E}[\hat{D}_{P+1,r}] - \mathbb{E}[\hat{D}_{P,r}] + \mathrm{Pr}(S_{P+1} > T_{P,r})(\frac{1}{p_0}-1).
\end{equation}
Since $\mathbb{E}[D_{1,r}] = \frac{1}{p_0}+\frac{1}{p_r}-1$ and $\mathbb{E}[\hat{D}_{1,r}] = \frac{1}{p_r}$, we have
\begin{equation}
\label{eqn:D_Pr_D_Pr_hat_relation2}
\mathbb{E}[D_{P,r}] = \mathbb{E}[\hat{D}_{P,r}] + (\frac{1}{p_0}-1)\left(1+\sum\nolimits_{j=1}^{P-1} \mathrm{Pr}(S_{j+1} > T_{j,r})\right)
\end{equation}
Moreover, due to \eqref{eqn:P_P_1}, \eqref{eqn:D_Pr_D_Pr_hat_relation1} and $\mathbb{E}[\hat{D}_{P+1,r}] - \mathbb{E}[\hat{D}_{P,r}] = \frac{1}{p_r}$, the following lemma can be obtained.

\begin{lemma}
\label{lemma:S_P+1_T_P_r}
When $p_0 < 1$, $\mathrm{Pr}(S_{P+1}\ > T_{P,r})=\frac{1}{1-p_0} + \frac{p_0}{1-p_0}\sum\nolimits_{i=0}^{P-1}B(i)$.
\end{lemma}

We next make a connection between $\mathbb{E}[D_{P}]$ and $\mathbb{E}[\hat{D}_{P}]$ similar to that between $\mathbb{E}[D_{P,r}]$ and $\mathbb{E}[\hat{D}_{P,r}]$ in \eqref{eqn:D_Pr_D_Pr_hat_relation2}. For $R$ independent geometrically distributed random variables $N_1, \ldots, N_R$ (starting from $1$) with respective parameters $p_1, \ldots, p_R$, define 
\begin{equation}
\label{eqn:E_max_definition}
E_{\max} = \mathbb{E}[\max\{N_1, \ldots, N_R\}].
\end{equation}
$E_{\max}$ can be explicitly computed by the min-max identity (see, e.g., \cite{SheldonRoss}). Based on $E_{\max}$,
\begin{equation}
\label{eqn:ED1_ED1_hat_relation}
\mathbb{E}[D_1] = 1/p_0 + E_{\max} - 1,~\mathbb{E}[\hat{D}_{1}] = E_{\max},~\mathbb{E}[D_1] = \mathbb{E}[\hat{D}_{1}] + 1/p_0 - 1.
\end{equation}%
For $P \geq 2$, the connection between $\mathbb{E}[D_P]$ and $\mathbb{E}[\hat{D}_P]$ can be analyzed by conditioning on the relation among $S_{P+1}$, $T_{P+1, r}$ and $T_{P, r}$, $1 \leq r \leq R$. To ease the following presentation, we first elaborate the case of $R = 2$, and then give a general conclusion for $R \geq 2$.

\begin{lemma}
\label{lemma:recursive_expression_of_fullduplex1}
For the case $R = 2$,
\begin{equation}
\begin{split}
\label{eqn:recursive_expression_of_fullduplex1}
& \mathbb{E}[D_{P+1}] - \mathbb{E}[D_{P}] \\
=& \mathrm{Pr}(T_{P+1,1} \leq T_{P,2})\frac{1}{p_2} + \mathrm{Pr}(T_{P+1,2} \leq T_{P,1})\frac{1}{p_1} +  \mathrm{Pr}(T_{P+1,1} > T_{P,2}, T_{P+1,2} > T_{P,1})E_{\max}+\\
 & \mathrm{Pr}(S_{P+1} > \max\{T_{P,1}, T_{P,2}\})(\frac{1}{p_0}-1).
\end{split}
\end{equation}
\begin{IEEEproof}
We analyze $\mathbb{E}[D_{P+1}] - \mathbb{E}[D_{P}]$ by conditioning on the following $4$ different cases:

\begin{itemize}
\item Let $A$ represent the case $T_{P+1,1} \leq T_{P,2}$ and $T_{P+1,2} > T_{P,1}$, which is equivalent to $T_{P+1,1} \leq T_{P,2}$. In this case, when both receivers have obtained $P$ packets, receiver $1$ has obtained $(P+1)^{st}$ packet as well. Thus, it only takes additional $1/p_2$ timeslots on average for receiver $2$ to get the $(P+1)^{st}$ packet, that is, $\mathbb{E}[D_{P+1} - D_{P}~|~A] = 1/p_2$.
\item Let $B$ represent the case $T_{P+1,2} \leq T_{P,1}$ and $T_{P+1,1} > T_{P,2}$, which is equivalent to $T_{P+1,2} \leq T_{P,1}$. In a similar argument to case $A$, we have $\mathbb{E}[D_{P+1} - D_{P} ~|~ B] = 1/p_1$.
\item Let $C$ represent the case $T_{P+1,2} > T_{P,1}$ and $T_{P+1,1} > T_{P,2}$. We further divide $C$ into two subcases, that is, $S_{P+1} > \max\{T_{P,1}, T_{P,2}\}$ and $S_{P+1} \leq \max\{T_{P,1}, T_{P,2}\}$. In the first subcase $S_{P+1} > \max\{T_{P,1}, T_{P,2}\}$, when both receivers have obtained $P$ packets, it takes extra $1/p_0$ timeslots on average for the RS to get the $(P+1)^{st}$ packet and then $E_{\max}-1$ timeslots on average to make both receivers obtain the $(P+1)^{st}$ packets, that is,
    \begin{equation*}
    \begin{split}
    &\mathbb{E}[D_{P+1} - D_P ~|~ C, S_{P+1} > \max\{T_{P,1}, T_{P,2}\}] =  1/p_0 + E_{\max} - 1.
    \end{split}
    \end{equation*}
     In the second subcase $S_{P+1} \leq \max\{T_{P,1}, T_{P,2}\}$, when both receivers have obtained $P$ packets, the RS has already received the $(P+1)^{st}$ packet, so it takes extra $E_{\max}$ timeslots on average to make both receivers obtain the $(P+1)^{st}$ packets, that is,
    \begin{equation*}
    \begin{split}
    \mathbb{E}[D_{P+1} - D_P ~|~ C, S_{P+1} \leq \max\{T_{P,1}, T_{P,2}\}]  = E_{\max}.
    \end{split}
    \end{equation*}
    In all,
    \begin{equation*}
    \begin{split}
    &\mathbb{E}[D_{P+1} - D_P ~|~ C] \\
    =& (1/p_0 + E_{\max}-1)\mathrm{Pr}(S_{P+1} > \max\{T_{P,1}, T_{P,2}\} ~|~ C) + E_{\max}\mathrm{Pr}(S_{P+1} \leq \max\{T_{P,1}, T_{P,2}\} ~|~ C) \\
    = &E_{\max} + (1/p_0-1)\mathrm{Pr}(S_{P+1} > \max\{T_{P,1}, T_{P,2}\} ~|~ C),
    \end{split}
    \end{equation*}
    and consequently
   \begin{equation*}
    \begin{split}
    &\mathbb{E}[D_{P+1} - D_P ~|~ C]\mathrm{Pr}(C) \\
    =&E_{\max}\mathrm{Pr}(C) + ~(1/p_0-1)\mathrm{Pr}(S_{P+1} > \max\{T_{P,1}, T_{P,2}\} ~|~ C)\mathrm{Pr}(C) \\
    =&E_{\max}\mathrm{Pr}(C) + (1/p_0-1)\mathrm{Pr}(S_{P+1} > \max\{T_{P,1}, T_{P,2}\})
    \end{split}
    \end{equation*}
\item The last case $T_{P+1,1} \leq T_{P,2}$ and $T_{P+1,2} \leq T_{P,1}$ has probability $0$ to occur.
\end{itemize}

Eq. \eqref{eqn:recursive_expression_of_fullduplex1} can now be proved to be correct due to $\mathbb{E}[D_{P+1} - D_{P}] = \mathbb{E}[D_{P+1} - D_P ~|~ A]\mathrm{Pr}(A) +\mathbb{E}[D_{P+1} - D_P ~|~ B]\mathrm{Pr}(B)+\mathbb{E}[D_{P+1} - D_P ~|~ C]\mathrm{Pr}(C)$.
\end{IEEEproof}
\end{lemma}

Let $\hat{T}_{j,r}$ denote the number of timeslots receiver $r$ takes to receive the $j^{th}$ packet in the special case $p_0 = 1$. Thus, \eqref{eqn:recursive_expression_of_fullduplex1} implies
\begin{equation}
\label{eqn:recursive_eq_of_wb}
\begin{split}
&\mathbb{E}[\hat{D}_{P+1}]-\mathbb{E}[\hat{D}_{P}] \\
=&  \mathrm{Pr}(\hat{T}_{P+1,1} \leq \hat{T}_{P,2})\frac{1}{p_2} + \mathrm{Pr}(\hat{T}_{P+1,2} \leq \hat{T}_{P,1})\frac{1}{p_1} + \mathrm{Pr}(\hat{T}_{P+1,1} > \hat{T}_{P,2}, \hat{T}_{P+1,2} > \hat{T}_{P,1})E_{\max}.
\end{split}
\end{equation}
Let $\varepsilon_{P}$ denote the following identity
\begin{equation}
\label{eqn:epsilon_identity}
\begin{split}
 \varepsilon_{P}
 = &(\mathrm{Pr}(T_{P,1} = T_{P,2}) - \mathrm{Pr}(\hat{T}_{P,1} = \hat{T}_{P,2})) \frac{(1-p_1)(1-p_2)(p_1+p_2)}{p_1p_2(p_1+p_2 - p_1p_2)} + \\
 &(\mathrm{Pr}(\hat{T}_{P,1} > \hat{T}_{P,2}) - \mathrm{Pr}(T_{P,1} > T_{P,2}))\frac{p_1(1-p_2)}{p_2(p_1+p_2 - p_1p_2)} + \\
 &(\mathrm{Pr}(\hat{T}_{P,2} > \hat{T}_{P,1}) - \mathrm{Pr}(T_{P,2} > T_{P,1}))\frac{(1-p_1)p_2}{p_1(p_1+p_2 - p_1p_2)}
 \end{split}
\end{equation}

\begin{theorem}
\label{thm:E[D_P]_recursive_formula}
For the case $R = 2$,
\begin{equation}
\label{eqn:equivalent_recursive_expression_of_E[D_P]}
 \mathbb{E}[D_{P+1}] - \mathbb{E}[D_{P}]
 = \mathbb{E}[\hat{D}_{P+1}] - \mathbb{E}[\hat{D}_{P}] + \varepsilon_{P+1} + \mathrm{Pr}(S_{P+1} > \max\{T_{P,1}, T_{P,2}\})(\frac{1}{p_0}-1).
 \end{equation}
\begin{IEEEproof}
Please refer to Appendix-\ref{Appendix:proof_of Theorem6}.
\end{IEEEproof}
\end{theorem}

Eq. \eqref{eqn:equivalent_recursive_expression_of_E[D_P]} in Theorem \ref{thm:E[D_P]_recursive_formula} extends \eqref{eqn:D_Pr_D_Pr_hat_relation1} from the single receiver case to the case $R = 2$. It implies an approximation, which can also serve as a lower bound, for $\mathbb{E}[D_{P}]$ based on $\mathbb{E}[\hat{D}_{P}]$. %
First, because $\mathrm{Pr}(S_{P+1} > T_{P, 1} | S_{P+1} > T_{P, 2}) \geq \mathrm{Pr}(S_{P+1} > T_{P, 1})$, we have
\begin{equation}
\label{eq:inequality_proof}
\begin{split}
&\mathrm{Pr}(S_{P+1} > \max\{T_{P, 1}, T_{P, 2}\}) \geq \mathrm{Pr}(S_{P+1} > T_{P, 1})\mathrm{Pr}(S_{P+1} > T_{P, 2}).
\end{split}
\end{equation}
Since $\mathrm{Pr}(S_{P+1} > T_{P, 1})$ and $\mathrm{Pr}(S_{P+1} > T_{P, 2})$ can be explicitly computed based on Lemma \ref{lemma:S_P+1_T_P_r} together with the recursive formula \eqref{eqn:B_i}, we shall adopt $\mathrm{Pr}(S_{P+1} > T_{P, 1})\mathrm{Pr}(S_{P+1} > T_{P, 2})$ as an explicitly computable lower bound on $\mathrm{Pr}(S_{P+1} > \max\{T_{P, 1}, T_{P, 2}\})$. %
Second, we shall omit $\varepsilon_{P+1}$ in the approximation whose performance will be justified below. %
For brevity, let $\tilde{D}_{P}$ denote
\begin{equation}
\label{eq:Tilde_D_P}
\tilde{D}_{P} = (\frac{1}{p_0}-1)\left(1+\sum\nolimits_{j=1}^{P-1} \mathrm{Pr}(S_{j+1} > T_{j, 1})\mathrm{Pr}(S_{j+1} > T_{j, 2})\right).
\end{equation}
\begin{theorem}
\label{thm:E_DP_approximation}
For the case $R = 2$ and $P \geq 2$,
\begin{align}
\label{eqn:E_DP_approximation}
\mathbb{E}[D_P] = &\mathbb{E}[\hat{D}_P] + \sum\nolimits_{j = 2}^{P} \varepsilon_{j} + (\frac{1}{p_0}-1)(1+\sum\nolimits_{j = 1}^{P-1} \mathrm{Pr}(S_{j+1} > \max\{T_{j,1}, T_{j,2}\})) \nonumber \\
\geq &\mathbb{E}[\hat{D}_P] + \tilde{D}_P.
\end{align}
\begin{IEEEproof}
The first equation is a direct consequence of \eqref{eqn:equivalent_recursive_expression_of_E[D_P]} and \eqref{eqn:ED1_ED1_hat_relation}. By \eqref{eq:inequality_proof},
\begin{equation*}
\mathbb{E}[D_{P}]
 \geq \mathbb{E}[\hat{D}_P] + \sum\nolimits_{j = 2}^{P} \varepsilon_{j} + \tilde{D}_P.
\end{equation*}
It remains to prove $\sum\nolimits_{j = 2}^{P} \varepsilon_{j} \geq 0$, which can be found in Appendix-\ref{Proof of Theorem_E_DP_approximation}.
\end{IEEEproof}
\end{theorem}

We can now consider $\mathbb{E}[\hat{D}_{P}] + \tilde{D}_{P}$ as an approximation as well as a lower bound for $\mathbb{E}[D_P]$ when $R =2$ and $P \geq 2$. Notice that $\mathbb{E}[\hat{D}_{P}]$ and $\tilde{D}_{P}$ can be explicitly computed by \eqref{eqn:perfect_for_wireless_broadcast} and by Lemma \ref{lemma:S_P+1_T_P_r} together with the recursive formula \eqref{eqn:B_i}, respectively. %

Before proceeding to generalize the approximation $\mathbb{E}[\hat{D}_{P}]+\tilde{D}_{P}$ of $\mathbb{E}[D_{P}]$ to the case $R > 2$, we briefly discuss the approximation accuracy, which depends on how close $\tilde{D}_P$ approximates the difference $\mathbb{E}[D_P] - \mathbb{E}[\hat{D}_{P}]$. %
In the process of obtaining the approximation value $\tilde{D}_{P}$ in Theorem \ref{thm:E_DP_approximation}, we neglect the term $\sum\nolimits_{j = 2}^P \varepsilon_{j} \geq 0$, approximate $\sum\nolimits_{j = 1}^{P-1} \mathrm{Pr}(S_{j+1} > \max\{T_{j,1}, T_{j,2}\})$ as
$\sum\nolimits_{j=1}^{P-1} \mathrm{Pr}(S_{j+1} > T_{j, 1})\mathrm{Pr}(S_{j+1} > T_{j, 2})$, and add $1 - 1/p_0$, which represents the expected number of timeslots the RS takes to obtain the first packet. %
Observe that with increasing $P$, both $\mathrm{Pr}(S_{P+1} > T_{P,1})$ and $\mathrm{Pr}(S_{P+1} > T_{P,2})$ decrease by Lemma \ref{lemma:S_P+1_T_P_r}, and so does $\mathrm{Pr}(S_{P+1} > \max\{T_{P,1}, T_{P,2}\})$. %
Without loss of generality, we next analyze the loss by approximating $\mathbb{E}[D_P] - \mathbb{E}[\hat{D}_P]$ as $\tilde{D}_P$ via the following $3$ cases.
\begin{itemize}
\item Case 1: $p_0 > \max\{p_1,p_2\}$, so that $\mathrm{Pr}(S_{P+1} > T_{P,1})$, $\mathrm{Pr}(S_{P+1} > T_{P,2})$ and $\mathrm{Pr}(S_{P+1} > \max\{T_{P,1}, T_{P,2}\})$ are small. Thus, $\mathrm{Pr}(S_{P+1} > \max\{T_{P,1}, T_{P,2}\}) -  \mathrm{Pr}(S_{P+1} > T_{P,1})\mathrm{Pr}(S_{P+1} > T_{P,2})$ tends to zero fast with increasing $P$. Moreover, $\varepsilon_P$ also converges to $0$ fast with increasing $P$. Actually, with increasing $j$, when a receiver obtains the $j^{th}$ new packet, there is a higher probability that the RS has obtained at least $j+1$ packets from the BS. It turns out that the approximation of the transmission scenario from the RS to receivers as a wireless broadcast is accurate, that is, $\tilde{D}_{P}/P$ is a close approximation of $(\mathbb{E}[D_P] - \mathbb{E}[\hat{D}_P])/P$.
\item Case 2: $p_1<p_0\leq p_2$, so that both $\mathrm{Pr}(S_{P+1} > T_{P,1})$ and $\mathrm{Pr}(S_{P+1} > \max\{T_{P,1}, T_{P,2}\})$ are relatively small. The gap between $\mathrm{Pr}(S_{P+1} > \max\{T_{P,1}, T_{P,2}\})$ and $\mathrm{Pr}(S_{P+1} > T_{P,1})\mathrm{Pr}(S_{P+1} > T_{P,2})$ is negligible for small $P$. However, for larger $P$, the approximation of $\mathrm{Pr}(S_{P+1} > \max\{T_{P,1}, T_{P,2}\})$ as $\mathrm{Pr}(S_{P+1} > T_{P,1})\mathrm{Pr}(S_{P+1} > T_{P,2})$ becomes less accurate than that in Case 1. %
    Moreover, $\varepsilon_P$ tends to $0$ with $P$ increasing (but not as fast as in Case 1). Hence, the approximation of $(\mathbb{E}[D_P] - \mathbb{E}[\hat{D}_P])/P$ by $\tilde{D}_P/P$ in this case performs a little less accurate than that in Case 1 (with the same $p_1$, $p_2$).
\item Case 3: $p_0 \leq \min\{p_1,p_2\}$. In this case, with increasing $P$, $\mathrm{Pr}(S_{P+1}> T_{P,1})$,
$\mathrm{Pr}(S_{P+1}> T_{P,2})$ and $\mathrm{Pr}(S_{P+1} > \max\{T_{P, 1}, T_{P, 2}\})$ decrease slower than those in Case 1 and 2 (which have the same $p_1$, $p_2$ but larger $p_0$). Consequently, $\mathrm{Pr}(S_{P+1} > \max\{T_{P, 1}, T_{P, 2}\})-\mathrm{Pr}(S_{P+1} > T_{P,1})\mathrm{Pr}(S_{P+1} > T_{P,2})$ is not negligible even for large $P$. Similarly, $\varepsilon_P$ does not converge to $0$ with $P$ increasing. As a result, for large $P$, the approximation of $(\mathbb{E}[D_P] - \mathbb{E}[\hat{D}_P])/P$ by $\tilde{D}_P/P$ is not as accurate as in Case 1 and 2 (with the same $p_1$, $p_2$).
\end{itemize}

Notice that for every receiver $r$, the expected completion delay $\mathbb{E}[D_{P,r}]$, which has been explicitly characterized in Theorem \ref{thm:perfect scheme with_buffer} and can be efficiently computed according to \eqref{eqn:P_P_1}, is naturally a lower bound of $\mathbb{E}[D_{P}]$. %
When $p_1$ is much smaller than $p_2$, or $p_1 \leq p_2$ with large enough $P$, the probability that the completion delay at receiver $2$ is no larger than that at receiver $1$, that is, $\mathrm{Pr}(D_{P,2} \leq D_{P,1})$ is high. %
As a result, the following simple lower bound for $\mathbb{E}[D_{P}]$
\begin{equation}
\label{eq:lower_bound2}
\mathbb{E}[D_{P}] \geq \max\{\mathbb{E}[D_{P,1}],\mathbb{E}[D_{P,2}]\}
\end{equation}
may provide a better approximation compared with \eqref{eqn:E_DP_approximation} when the difference of $p_1$ and $p_2$ is large or $P$ is relatively large, particularly for Case 2 and 3 discussed above.
\begin{corollary}
\label{corollary:lower_bound_R2}
For the case $R = 2$ and $P\geq 2$,
\begin{equation}
\label{eqn:lower_bound_complete_R2}
\begin{split}
&\mathbb{E}[D_P] \geq \max\{\mathbb{E}[D_{P,1}], \mathbb{E}[D_{P,2}], \mathbb{E}[\hat{D}_P]+ \tilde{D}_{P}\}.
\end{split}
\end{equation}
\end{corollary}

We now generalize the approximation of $\mathbb{E}[D_P]$ as $\mathbb{E}[\hat{D}_{P}]+\tilde{D}_{P}$ from $R = 2$ to $R \geq 2$. %
Write
\begin{equation}
\label{eqn:recursive_eq_of_D_P_hat_general_R}
\hat{\Psi}_{P+1} = \mathbb{E}[\hat{D}_{P+1}] - \mathbb{E}[\hat{D}_{P}].
\end{equation}
Because $\mathbb{E}[D_{P+1} - D_{P} | S_{P+1} > \max\nolimits_{1\leq r \leq R}T_{P,r}] = \frac{1}{p_0}-1$, we can express
\begin{equation}
\label{eqn:recursive_eq_of_D_P_general_R}
\mathbb{E}[D_{P+1}] - \mathbb{E}[D_{P}]
= \Psi_{P+1} + \mathrm{Pr}(S_{P+1} > \max\nolimits_{1\leq r \leq R}T_{P,r})(\frac{1}{p_0}-1),
\end{equation}
where $\Psi_{P+1} = \mathbb{E}[D_{P+1} - D_{P} | S_{P+1} \leq \max_{1\leq r \leq R}T_{P,r}]\mathrm{Pr}(S_{P+1} \leq \max_{1\leq r \leq R}T_{P,r})$. Based on \eqref{eqn:recursive_eq_of_D_P_hat_general_R}, \eqref{eqn:recursive_eq_of_D_P_general_R} and \eqref{eqn:ED1_ED1_hat_relation}, we have
\begin{equation}
\label{eq:E_D_P_mul_R}
\mathbb{E}[D_P] = \mathbb{E}[\hat{D}_P] + \sum\nolimits_{j=2}^P(\Psi_{j}-\hat{\Psi}_j) +  (\frac{1}{p_0}-1)(1+\sum\nolimits_{j=1}^{P-1}\mathrm{Pr}(S_{j+1} > \max\nolimits_{1\leq r \leq R}T_{j,r}))
\end{equation}

For $R = 2$, it has been proved in Theorem \ref{thm:E_DP_approximation} that $\sum\nolimits_{j=2}^P(\Psi_{j}-\hat{\Psi}_j) = \sum\nolimits_{j = 2}^{P} \varepsilon_{j} \geq 0$. For $R > 2$, further analysis of $\sum\nolimits_{j=2}^P(\Psi_{j}-\hat{\Psi}_j)$ involves much more details which are beyond the scope of this paper, so we directly ignore this term in our approximation. Because Similar to \eqref{eq:inequality_proof}, $\mathrm{Pr}(S_{P+1} > \max_{1\leq r \leq R}T_{P,r}) \geq \prod_{1\leq r \leq R} \mathrm{Pr}(S_{P+1} > T_{P, r})$. Consequently, we obtain the following approximation of $\mathbb{E}[D_P]$
\begin{equation}
\label{eqn:EDP_aproximation_general_R}
\mathbb{E}[D_P] \approx \mathbb{E}[\hat{D}_P] + (\frac{1}{p_0}-1)(1+\sum\nolimits_{j=1}^{P-1}\mathrm{Pr}(S_{j+1} > \max\nolimits_{1\leq r \leq R}T_{j,r})) \geq \mathbb{E}[\hat{D}_P] + \tilde{D}_P,
\end{equation}
where $\tilde{D}_P$ is generalized from \eqref{eq:Tilde_D_P} to denote $\tilde{D}_{P} = (\frac{1}{p_0}-1)\left(1+\sum_{j=1}^{P-1}\prod_{r = 1}^R \mathrm{Pr}(S_{j+1} > T_{j, r})\right)$.
Recall that $\mathbb{E}[\hat{D}_P]$ represents the expected system completion delay for the special case $p_0 = 1$, which degenerates to the wireless broadcast, and according to \eqref{eqn:perfect_for_wireless_broadcast}, we have the following explicit expression
\begin{equation}
\label{eqn:perfect_for_wireless_broadcast_explicit}
\mathbb{E}[\hat{D}_P] = P+\sum\nolimits_{d \geq 0} \left(1 - \prod\nolimits_{1 \leq r \leq R} \sum\nolimits_{j=0}^{d} \binom{P+j-1}{P-1} p_{r}^P(1-p_r)^{j}\right).
\end{equation}
Based on Lemma \ref{lemma:S_P+1_T_P_r} and the definition of $B(i)$ in \eqref{eqn:Bi_definition}, we can also express $\tilde{D}_{P}$ in the following closed-form, which is a function of $p_0$ and $p_r$
\begin{equation}
\label{eqn:tilde_D_P_explicit}
\begin{split}
\tilde{D}_{P} = &(\frac{1}{p_0}-1)\left(1+\sum_{j=1}^{P-1}\prod_{r = 1}^R \mathrm{Pr}(S_{j+1} > T_{j, r})\right) \\
= &(\frac{1}{p_0}-1)\left(1+\sum_{j=1}^{P-1}\prod_{r = 1}^R (\frac{1}{1-p_0} + \frac{p_0}{1-p_0}\sum\nolimits_{i=0}^{j-1}\sum\nolimits_{k = 0}^{i} \frac{\frac{1}{k+1}\binom{i+k}{i}\binom{i}{k}(p_0p_r)^i}{(p_0p_r - p_0 - p_r)^{i+k+1}})\right).
\end{split}
\end{equation}
Moreover, notice that $\tilde{D}_{P}$ can be explicitly computed by the following recursive procedure:
\begin{itemize}
\item $\tilde{D}_1 = \mathbb{E}[D_1] - \mathbb{E}[\hat{D}_1] = 1/p_0  - 1$.
\item For $P \geq 1$, $\tilde{D}_{P+1} = \tilde{D}_{P} + \prod_{r=1}^{R}\mathrm{Pr}(S_{P+1}>T_{P,r})(\frac{1}{p_0}-1)
 = \tilde{D}_{P} + \prod_{r=1}^{R}(\frac{1}{p_0}+ \sum_{i=0}^{P-1}B_r(i))$, where $B_r(i)$ denotes $B(i)$ in \eqref{eqn:Bi_definition} for given $p_r$ and it can be recursively computed by \eqref{eqn:B_i}.
\end{itemize}

For fixed $p_r$, $\mathbb{E}[\hat{D}_P]$ keeps the same so that the accuracy for the approximation $\mathbb{E}[\hat{D}_P] + \tilde{D}_P$ of $\mathbb{E}[D_P]$ depends on $\tilde{D}_P$. %
In particular, both $\frac{1}{p_0}-1$ and $\mathrm{Pr}(S_{P+1} > T_{P, r})$ in $\tilde{D}_{P}$ decrease with increasing $p_0$, and $\tilde{D}_P$ converges to $0$ when $p_0$ increases to $1$, which represents the case $\mathbb{E}[D_P] = \mathbb{E}[\hat{D}_P]$. %
Moreover, as $\mathrm{Pr}(S_{P+1} > T_{P, r})$ can be explicitly computed by Lemma \ref{lemma:S_P+1_T_P_r}, the use of $\prod_{r = 1}^R \mathrm{Pr}(S_{P+1} > T_{P, r})$ in $\tilde{D}_P$ is to approximate the joint probability $\mathrm{Pr}(S_{P+1} > \max_{1\leq r \leq R}T_{P,r})$, which is a key component in characterizing $\mathbb{E}[D_P]$ in \eqref{eq:E_D_P_mul_R}. Analogous to the discussion above Corollary \ref{corollary:lower_bound_R2} for the case $R=2$, we further argue that for sufficiently large $P$, when $p_0 > \max\nolimits_{1\leq r \leq R}p_r$, $\prod_{r = 1}^R \mathrm{Pr}(S_{P+1} > T_{P, r})$ is a closer approximation of $\mathrm{Pr}(S_{P+1} > \max_{1\leq r \leq R}T_{P,r})$ compared with the case $p_0 \leq \max_{1\leq r \leq R}p_r$, while the approximation performs the worst for the case $p_0 \leq  \min\nolimits_{1\leq r \leq R}p_r$.

In \eqref{eqn:EDP_aproximation_general_R}, we use the sign ``$\approx$'' instead of ``$\geq$'' because we did not prove the term $\sum\nolimits_{j=2}^P(\Psi_{j}-\hat{\Psi}_j)$ in the explicit characterization of $\mathbb{E}[D_P]$ in \eqref{eq:E_D_P_mul_R} to be non-negative for $R > 2$, which we conjecture to be correct. %
By further taking the simple lower bound $\mathbb{E}[D_{P}] \geq \max\nolimits_{1\leq r \leq R}\mathbb{E}[D_{P,r}]$ into account,
we obtain the following approximation of $\mathbb{E}[D_P]$ for $R \geq 2$,
\begin{equation}
\label{eqn:D_P+1_mul_receiver}
\mathbb{E}[D_{P}] \gtrsim \max\{\max\nolimits_{1\leq r \leq R}\mathbb{E}[D_{P,r}], \mathbb{E}[\hat{D}_P]+\tilde{D}_{P}\},
\end{equation}
where $\gtrsim$ is rigorously proved to be $\geq$ in Theorem \ref{thm:E_DP_approximation} and Corollary \ref{corollary:lower_bound_R2} for $R = 2$. The tightness of the above approximation will be numerically analyzed in the next section.

It is worthwhile noting that in the case that $R$ is relatively small and there is a receiver whose successful receiving probability $p_r$ is much smaller than others', $\mathbb{E}[D_{P,r}]$ is a better lower bound of $\mathbb{E}[D_{P}]$ compared with $\mathbb{E}[\hat{D}_P]+\tilde{D}_{P}$. Otherwise, $\mathbb{E}[\hat{D}_P]+\tilde{D}_{P}$ will be a better approximation of $\mathbb{E}[D_{P}]$, because it is built upon the expected system completion delay of wireless broadcast, which has already taken all receivers into consideration. %
In the full-duplex relay network modeled in this paper, as $\mathbb{E}[D_{P}]$ is a lower bound for the expected system completion delay of an arbitrary RLNC scheme, so is $\max\{\max\nolimits_{1\leq r \leq R}\mathbb{E}[D_{P,r}], \mathbb{E}[\hat{D}_P]+\tilde{D}_{P}\}$.

\section{Numerical Analysis}
In this section, we numerically analyze the average completion delay of the two fundamental perfect RLNC schemes and compare the numerical results with the theoretical characterizations obtained in the previous section. Moreover, we compare the performance of the two fundamental schemes with the one proposed in \cite{Chen&Meng2020} --- from the perspective of the average completion delay and the average buffer size taken at the RS, both of which are normalized by $P$. %
We adopt the following abbreviations in the figure legends. The FBPF perfect RLNC scheme, the perfect RLNC scheme without buffer, and the perfect RLNC scheme with buffer are respectively labeled as ``FBPF'', ``PRLNC w/t buffer'', and ``PRLNC w/ buffer''. The results obtained by simulation are labeled with ``simu'', and the theoretical results from Proposition \ref{thm:perfect scheme without_buffer} and Theorem \ref{thm:perfect scheme with_buffer} are labeled with ``th''. The performance for the single receiver and for the system is respectively labeled as ``single r'' and ``system''.
%
%
%
\begin{figure}
\centering
\hspace{-0.1in}
\subfigure[]
{\includegraphics[trim=10 2 13 17, clip,width=0.52\linewidth]{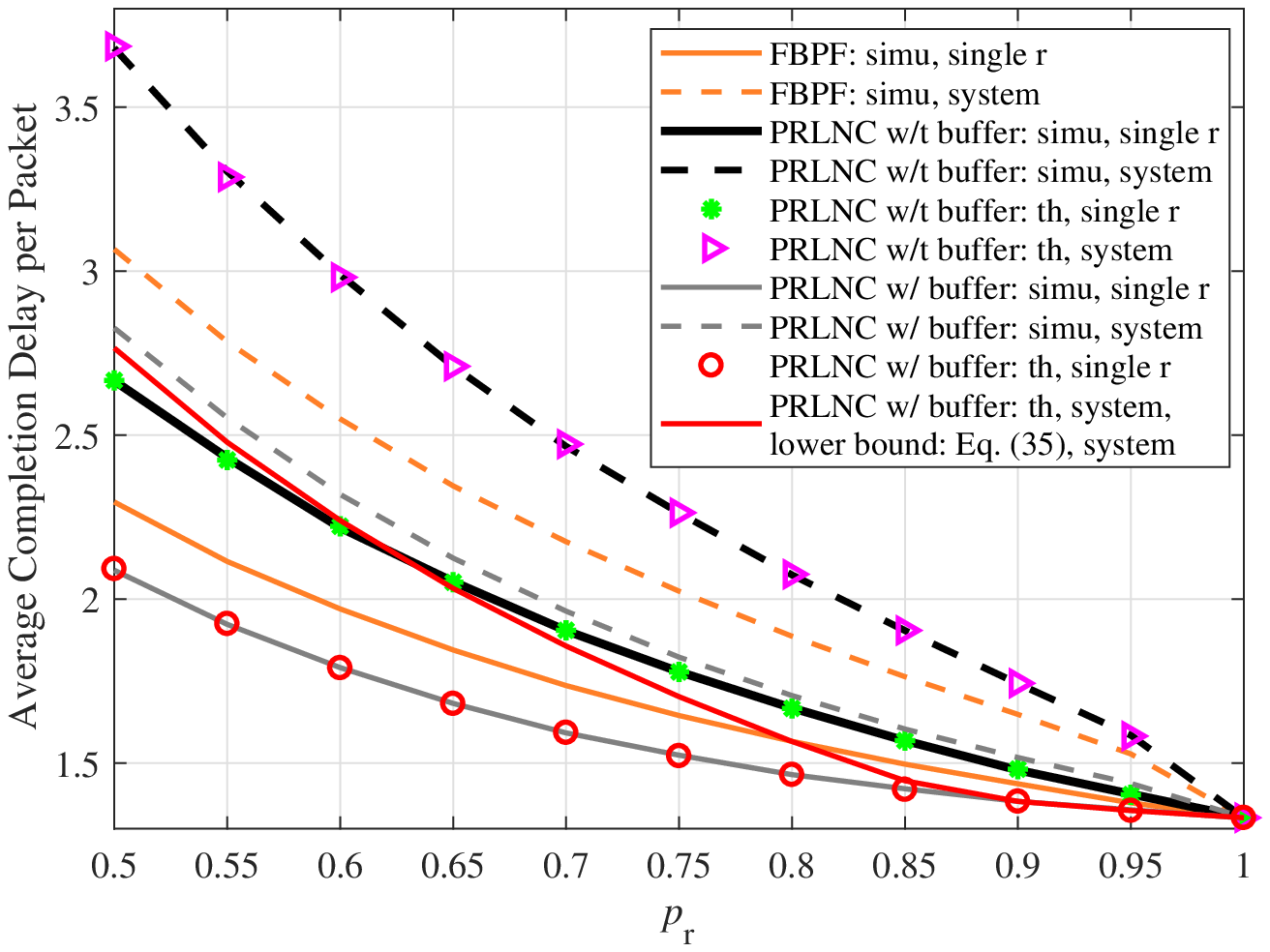}\label{fig:Completion delay per packet with varying pr}}
\hspace{-20pt}
\subfigure[]
{\includegraphics[trim=10 2 16 15, clip,width=0.52\linewidth]{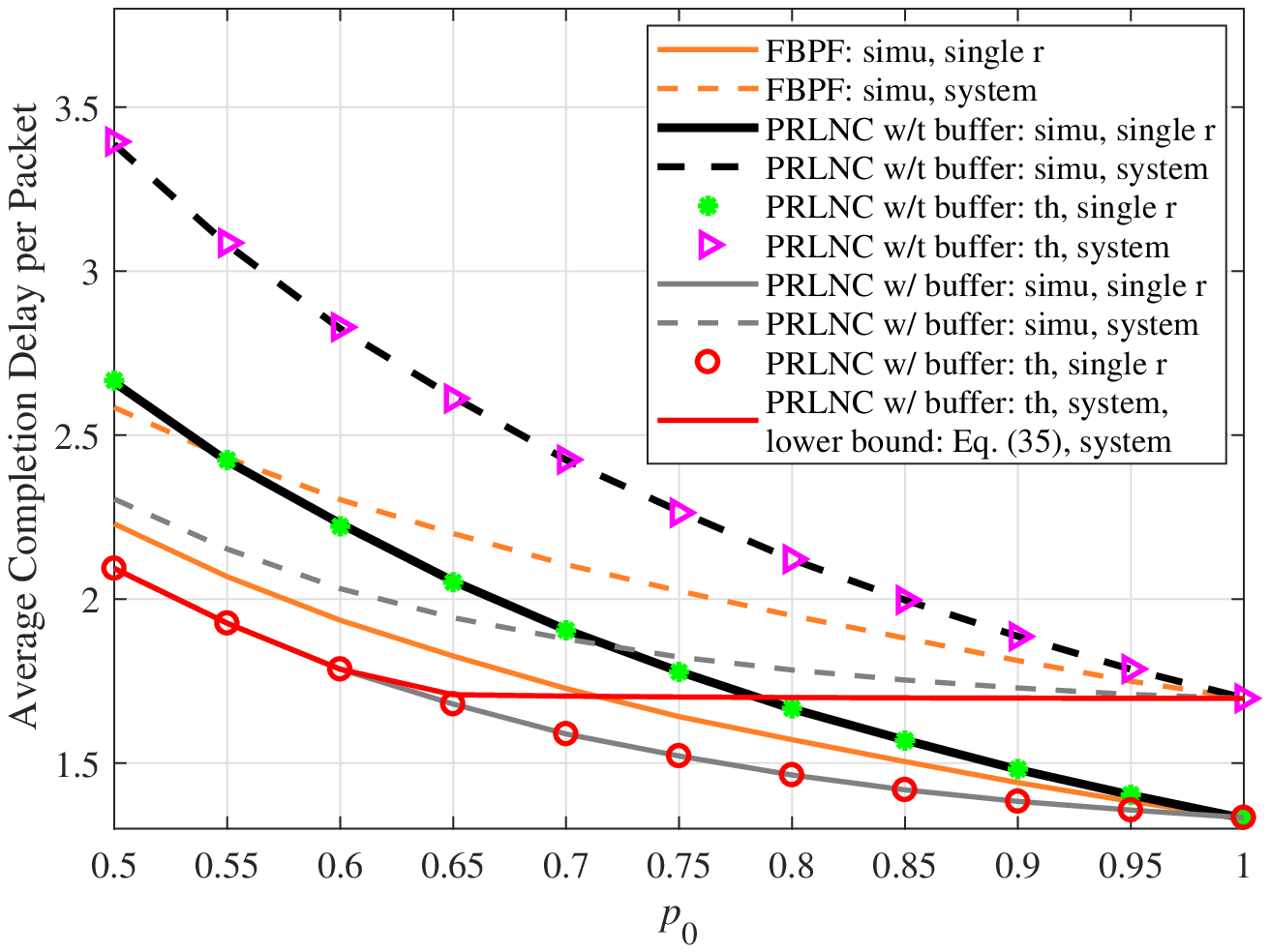}\label{fig:Completion delay per packet with varying p0}}
\caption{Average completion delay per packet of different schemes with: (a) fixed $P = 10$, $R = 10$, $p_0=0.75$ and varying $p_r$; (b) fixed $P = 10$, $R = 10$, $p_r=0.75$ and varying $p_0$.}
\label{fig:Completion delay per packet_plot}
\end{figure}
Under the settings $P = 10$, $R = 10$ and $p_0 = 0.75$, Fig. \ref{fig:Completion delay per packet with varying pr} depicts the average completion delay per packet at a single receiver as well as for the system with varying $p_r$. One may observe the followings. First, the average completion delay of every scheme decreases with increasing $p_r$, and converges to $1/p_0 = 1.33$. %
Second, the average completion delay of FBPF is upper bounded by that of perfect RLNC without buffer and lower bounded by that of perfect RLNC with buffer. Most importantly, the simulation results numerically verify the theoretical derivations in Proposition \ref{thm:perfect scheme without_buffer} and Theorem \ref{thm:perfect scheme with_buffer}, %
as well as validate the lower bound \eqref{eqn:D_P+1_mul_receiver} for the expected system completion delay for perfect RLNC with buffer.

Under the settings $P = 10$, $R = 10$ and $p_r = 0.75$, Fig. \ref{fig:Completion delay per packet with varying p0} depicts the average completion delay per packet at a single receiver as well as for the system with increasing $p_0$. In addition to similar observations to Fig. \ref{fig:Completion delay per packet with varying pr}, one may further conclude the followings. The average completion delay for the system of all three schemes converges to $1 + \frac{1}{P}\sum\nolimits_{d \geq 0} (1 - \prod\nolimits_{1 \leq r \leq R} I_{p_r}(P, d+1)) = 1.69$. %
Moreover, the plots for the average completion delay at a single receiver in Fig. \ref{fig:Completion delay per packet with varying pr} and Fig. \ref{fig:Completion delay per packet with varying p0} are almost identical, which infers that for all three schemes, the exchange of the values of $p_0$ and $p_r$ does not affect the completion delay performance at a single receiver. Theoretically, \eqref{eq:delay without_buffer_single_receiver} and \eqref{eqn:thm_w_buffer} justify this observation for perfect RLNC without buffer and with buffer, respectively. Lastly, with increasing $p_0$, the lower bound \eqref{eqn:D_P+1_mul_receiver} becomes tighter, which is in line with the discussion in the previous section.

Table \ref{table:Required buffer} lists the average buffer size per packet needed at the RS for the FBPF scheme with the settings $P = 10$, $R = 10$, $p_r = 0.75$ and different $p_0$. It is interesting to notice that the required buffer size increases with increasing $p_0$. In comparison, the perfect RLNC scheme without buffer demands no buffer, and the buffer size per packet of the perfect RLNC scheme with buffer is always $1$, which is $34\%$ smaller than that of the FBPF scheme when $p_0 = p_r = 0.75$.

\begin{figure}[t!]
\centering
\scalebox{0.7}
{\includegraphics[trim=10 5 10 23, clip]{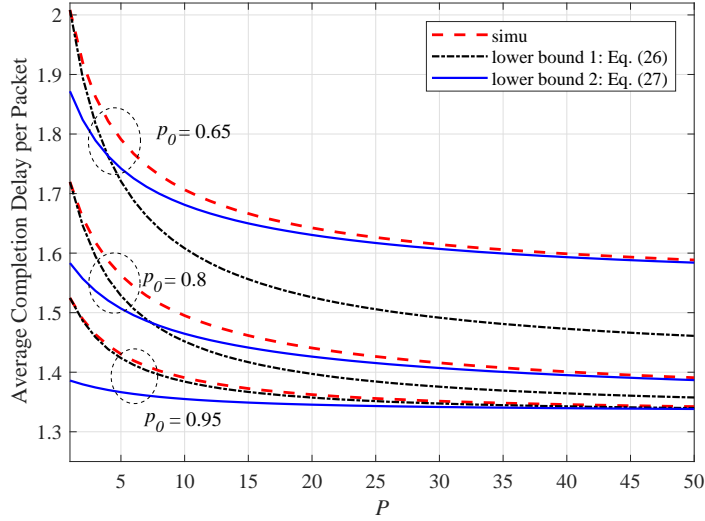}} 
\caption{Average system completion delay per packet and its lower bounds for perfect RLNC with buffer with fixed $R = 2, p_1 = 0.75, p_2 = 0.85$ and different $p_0$, $P$.}
\label{fig:Completion delay per packet R=2}
\end{figure}

\newcommand{\tabincell}[2]{\begin{tabular}{@{}#1@{}}#2\end{tabular}}
\begin{table}
\center
\caption{The average buffer size per packet of the FBPF scheme with fixed $P = 10$, $R = 10$, $p_r = 0.75$ and varying $p_0$}
\label{table:Required buffer}
\begin{tabular}{|c|c|c|c|c|c|c|}
  \hline
  $p_0$ & 0.5 & 0.6 & 0.7 &0.8 &0.9&1  \\
  \hline
  \tabincell{c}{Buffer Size}  & 1.293 &1.383	&1.474&	1.559&	1.63 &1.698  \\
  \hline
\end{tabular}
\end{table}

Fig. \ref{fig:Completion delay per packet_plot} has demonstrated the tightness of the lower bound \eqref{eqn:D_P+1_mul_receiver} for the expected system completion delay $\mathbb{E}[D_P]$ for perfect RLNC with buffer. Recall that the bound \eqref{eqn:D_P+1_mul_receiver} consists of two parts, both of which can be explicitly computed. One is $(\tilde{D}_{P}+\mathbb{E}[\hat{D}_P])/P$, which is a lower bound deduced from the perspective of wireless broadcast, and the other is $\max_{1\leq r\leq R}\mathbb{E}[D_{P,r}]/P$, which is the expected completion delay at the single receiver with the worst channel condition. %
Above Corollary \ref{corollary:lower_bound_R2}, we have discussed, for $R = 2$, the accuracy of the lower bound $(\tilde{D}_{P}+\mathbb{E}[\hat{D}_P])/P$ and conclude that $\max_{1\leq r\leq R}\mathbb{E}[D_{P,r}]/P$ is necessary to form the better lower bound \eqref{eqn:lower_bound_complete_R2}. In the remaining part of this section, we shall further numerically compare the accuracy of the two lower bounds $(\tilde{D}_{P}+\mathbb{E}[\hat{D}_P])/P$ and $\max_{1\leq r\leq R}\mathbb{E}[D_{P,r}]/P$.

Under the settings $R = 2$, $p_1 = 0.75, p_2 = 0.85$, Fig. \ref{fig:Completion delay per packet R=2} compares the average system completion delay per packet with the two lower bounds $(\tilde{D}_{P}+\mathbb{E}[\hat{D}_P])/P$ (labeled as ``lower bound $1$: Eq.  \eqref{eqn:E_DP_approximation}'') and $\mathbb{E}[D_{P,1}]/P$ (labeled as ``lower bound $2$: Eq. \eqref{eq:lower_bound2}''). The comparison is conducted under $3$ different choices of $p_0$, that is, $p_0 \in \{0.95,0.8,0.65\}$. 
We can conclude the following observations from Fig. \ref{fig:Completion delay per packet R=2} about the two lower bounds.
\begin{itemize}
\item First, the average system completion delay per packet as well as its two lower bounds decrease with fixed $P$ and increasing $p_0$. They also decrease and converge to some values with fixed $p_0$ and increasing $P$.
\item Second, for all $3$ choices of $p_0$, $(\tilde{D}_{P}+\mathbb{E}[\hat{D}_P])/P$ is tighter than $\mathbb{E}[D_{P,1}]/P$ for small $P$. This is because $(\tilde{D}_{P}+\mathbb{E}[\hat{D}_P])/P$ is obtained from the perspective of wireless broadcast so it takes all receivers' completion delay into account. In particular, the approximation of $\mathbb{E}[D_P] - \mathbb{E}[\hat{D}_P]$ by $\tilde{D}_{P}$ is relatively accurate for small $P$ and when $P = 1$, the bound $(\tilde{D}_{P}+\mathbb{E}[\hat{D}_P])$ is exactly equal to $\mathbb{E}[D_P]$, that is, $\tilde{D}_{P} = \mathbb{E}[D_P] - \mathbb{E}[\hat{D}_P] = 0$. %
\item Moreover, in the two cases with $p_0 > \max\{p_1, p_2\}$, $(\tilde{D}_{P}+\mathbb{E}[\hat{D}_P])/P$ is always better than $\mathbb{E}[D_{P,1}]/P$. %
    When $p_0 = 0.95$, $(\tilde{D}_{P}+\mathbb{E}[\hat{D}_P])/P$ converges to the average system completion delay $\mathbb{E}[D_P]$ with increasing $P$. However, when $p_0 \leq \max\{p_1, p_2\}$, $(\tilde{D}_{P}+\mathbb{E}[\hat{D}_P])/P$ decreases faster with increasing $P$, so that $\mathbb{E}[D_{P,1}]/P$ outperforms. This justifies the usefulness to supplement $\mathbb{E}[D_{P,1}]/P$ in the tighter lower bound \eqref{eqn:lower_bound_complete_R2}. %
\item Last, by comparing the three curves related to $(\tilde{D}_{P}+\mathbb{E}[\hat{D}_P])/P$, we can find that approximating $\mathbb{E}[D_P]/P$ by merely $\mathbb{E}[\hat{D}_P]/P$ will be much looser with decreasing $p_0$ because it does not take $p_0$ into consideration. Therefore, the additional term $\tilde{D}_{P}$ we introduce in the lower bound \eqref{eqn:E_DP_approximation} is indispensable in estimating $\mathbb{E}[D_P]$. %
\end{itemize}
In summary, what have been observed from Fig. \ref{fig:Completion delay per packet R=2} are consistent with the (3-case) discussion about the accuracy to approximate $\mathbb{E}[D_P] - \mathbb{E}[\hat{D}_P]$ as $\tilde{D}_{P}$ in the previous section (above Corollary \ref{corollary:lower_bound_R2}), and they affirm that \eqref{eqn:lower_bound_complete_R2} is a tighter lower bound than the individual use of \eqref{eqn:E_DP_approximation} or \eqref{eq:lower_bound2}.

\begin{figure}[t!]
\centering
\scalebox{0.7}
{\includegraphics[trim=10 5 10 23, clip]{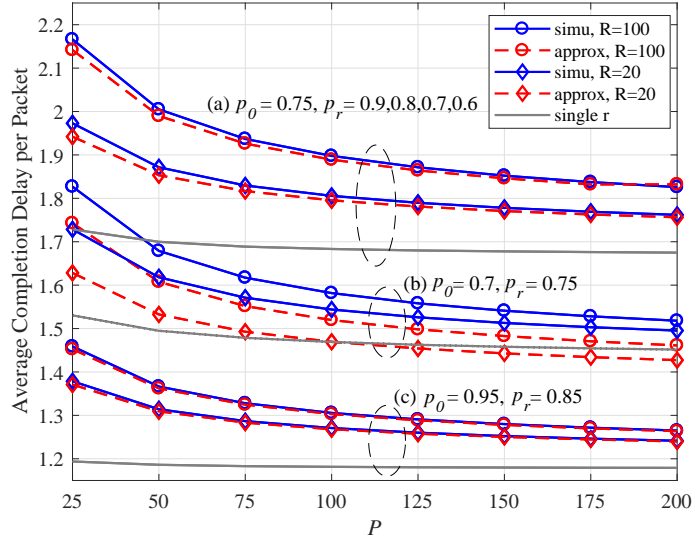}} 
\caption{Average system completion delay per packet and its lower bounds for perfect RLNC with buffer.}
\label{fig:Completion delay per packet mul_R}
\end{figure}

For multi-receiver case $R \in \{20,100\}$, Fig. \ref{fig:Completion delay per packet mul_R} compares the average system completion delay per packet $\mathbb{E}[D_P]/P$ with $(\tilde{D}_{P}+\mathbb{E}[\hat{D}_P])/P$ (labeled as ``approx'') and $\max\nolimits_{1\leq r \leq R}\mathbb{E}(D_{P,r})/P$ (labeled as ``single r''), which constitute the lower bound in \eqref{eqn:D_P+1_mul_receiver} as an extension of \eqref{eqn:lower_bound_complete_R2}. %
The comparison is conducted under three different settings of $p_0$ and $p_r$: (a) $p_0 = 0.75$ and $p_r$ is evenly distributed over $\{0.9, 0.8, 0.7, 0.6\}$; (b) $p_0 = 0.7$ and $p_r = 0.75$ for each receiver $r$; (c) $p_0 = 0.95$ and $p_r = 0.9$ for each receiver $r$. The followings can be observed.%
\begin{itemize}
\item For all $3$ settings, there is a noticeable gap between $\max\nolimits_{1\leq r \leq R}\mathbb{E}(D_{P,r})/P$ and the average system completion delay per packet $\mathbb{E}[D_P]/P$ even for large $P$ (compare with Fig. \ref{fig:Completion delay per packet R=2} in which the gap is very small for $P = 50$). This is mainly because the number of receivers considered herein is much more than that in Fig. \ref{fig:Completion delay per packet R=2}, so that the approximation accuracy from the perspective of a single receiver declines.
\item  In Settings (a) and (c), the curve of $(\tilde{D}_{P}+\mathbb{E}[\hat{D}_P])/P$ is close to the average system completion delay per packet $\mathbb{E}[D_P]/P$ for all $2$ choices of $R$. However, the tightness of $(\tilde{D}_{P}+\mathbb{E}[\hat{D}_P])/P$ in Setting (a) is slightly worse than that in Setting (c) mainly because half of the receivers' $p_r$ in Setting (a) are larger than $p_0$.
\item In Setting (b), the accuracy of $(\tilde{D}_{P}+\mathbb{E}[\hat{D}_P])/P$ to approximate $\mathbb{E}[D_{P}]/P$ is not as good as that in Settings (a) and (c), because $p_0 < p_r$ herein. Meanwhile, the approximation $\max\nolimits_{1\leq r \leq R}\mathbb{E}[D_{P,r}]/P$ becomes more accurate for large $P$, affirming that \eqref{eqn:D_P+1_mul_receiver} is a tighter lower bound of $\mathbb{E}[D_P]$ compared with $(\tilde{D}_{P}+\mathbb{E}[\hat{D}_P])/P$ or $\max\nolimits_{1\leq r \leq R}\mathbb{E}[D_{P,r}]/P$.
\end{itemize}
\section{Concluding Remarks}
In this paper, for full-duplex relay networks, we consider two fundamental perfect RLNC schemes to investigate the fundamental benefit RLNC can provide, and derive closed-form formulae for their expected completion delay, both at a single receiver and for the whole system. The expected completion delay of the two schemes can respectively serve as an upper bound for all perfect RLNC schemes and a lower bound for all RLNC schemes. It provides a theoretical guideline for future works on the detailed design of RLNC-based transmission schemes in the full-duplex relay networks. As an ensuing work, by adapting recently proposed efficient RLNC schemes such as Fulcrum \cite{Fulcrum2018} or circular-shift RLNC \cite{Sun_TIT19}, we will further design practical RLNC schemes with small buffer as well as low coding complexity at the RS, and with the completion delay performance closer to the theoretical limit characterized in this paper. Another ensuing work is to study the completion delay performance of perfect RLNC in a more general network model which contains direct links between the BS and the receivers as well as multiple full-duplex relays.
\appendix
\subsection{Proof of Eq. \eqref{eqn:expectation_w_buffer_final} }
\label{proof_of eq41}
For brevity, write $\Delta = p_0+p_r - p_0p_r = 1 - (1 - p_0)(1 - p_r)$. Let $q_{ij,i'j'}$ denote the total probability to enter state $(i',j')$ starting from state $(i,j)$ in the Markov chain under the constraint that the transitions from $(i,j)$ to $(i',j')$ are not allowed to visit states $(i + l, j + l')$ for all $0 \leq l < i' - i$ and $l' > l$. E.g., $q_{10,21} = \frac{p_0p_r}{\Delta^2}$ when $P \geq 3$. For brevity, $q_{00,ij}$ will be written as $q_{ij}$.

To reflect the number of original packets in the notation, write $E_P = \mathbb{E}[D_{P,r}]$. %
Stemming from \eqref{eqn:expectation_w_buffer}, we can obtain the following equivalent characterization
\begin{equation}
\label{eqn:expectation_w_buffer_2}
E_P =(1,0,\ldots,0)\sum\nolimits_{k\geq 0}\mathbf{P}^k \mathbf{1} = \sum\nolimits_{0\leq j \leq i \leq P} \frac{1}{1 - p_{ij,ij}} q_{ij}.
\end{equation}
Let $q_{ij}'$ represent the total probability that state $(i,j)$ can be entered starting from $(0,0)$ in the Markov chain $\mathcal{M}_{P-1}$, so that $q_{ij} = q_{ij}'$ for all $0 \leq j \leq i \leq P - 2$. For $j \leq P-2$, because $q_{jj}p_r + q_{Pj} + \frac{p_r}{\Delta}\sum_{i = j+1}^{P-1} q_{ij} = 1$ and $q_{jj}'p_r + q_{(P-1)j}' + \frac{p_r}{\Delta}\sum_{i = j+1}^{P-2} q_{ij}' = 1$, we have $q_{Pj} + \frac{p_1}{\Delta}q_{(P-1)j} = q_{(P-1)j}'$ and thus
\begin{equation}
\label{eqn:expectation_w_buffer_recursive_q}
\frac{q_{Pj}}{p_r}
= \frac{q_{(P-1)j}'}{p_r} - \frac{q_{(P-1)j}}{\Delta}.
\end{equation}
Since $1 - p_r$ is the $1$-step transition probability from state $(P, j)$ to itself in $\mathcal{M}_P$ as well as from state $(P-1,j)$ to itself in $\mathcal{M}_{P-1}$, and $1 - \Delta$ is the $1$-step transition probability from state $(P-1, j)$ to itself in $\mathcal{M}_P$, based on \eqref{eqn:expectation_w_buffer_2} and \eqref{eqn:expectation_w_buffer_recursive_q}, we obtain the following recursive expression
\begin{equation}
E_P = E_{P-1} + \frac{q_{(P-1)(P-1)}}{p_0} +  \frac{q_{P(P-1)}}{p_r}.
\end{equation}
As $1 = q_{PP} = q_{(P-1)(P-1)}p_r + q_{P(P-1)}$, we have $q_{P(P-1)} = 1 - q_{(P-1)(P-1)}p_r$ and
\begin{equation}
\label{eqn:expectation_w_buffer_recursive_whole}
E_P = E_{P-1} + \frac{1}{p_r} + \frac{1-p_0}{p_0}q_{(P-1)(P-1)}.
\end{equation}
As $E_1 = 1/p_0 + 1/p_r - 1$, in order to prove \eqref{eqn:thm_w_buffer} based on \eqref{eqn:expectation_w_buffer_recursive_whole} for $P \geq 2$, it remains to show
\begin{equation}
\label{eqn:expectation_w_buffer_final}
\frac{(1-p_0)q_{(P-1)(P-1)} -1}{p_0} = \sum\nolimits_{i=0}^{P-2}\sum\nolimits_{j=0}^{i} \frac{T_{i,j}(p_0p_r)^i}{(-\Delta)^{i+j+1}},
\end{equation}
which is equivalent to prove
\begin{equation}
\label{eqn:appendix_equivalent_condition}
\frac{1-p_0}{p_0}(q_{(P-1)(P-1)} - q_{(P-2)(P-2)}) = \sum\nolimits_{j=0}^{P-2} \frac{T_{P-2,j}(p_0p_r)^{P-2}}{(-\Delta)^{P+j-1}}
\end{equation}

In the remaining proof, we shall first make a connection between $q_{ii}$ and $q_{10,(i+1)i}$ for $0 < i < P$. Notice that for every $0 \leq j < i < P$, the $1$-step transition probability for state $(i, j)$ keeps the same, and $q_{ij,(i+1)j} = \frac{p_0}{\Delta}q_{i'i',(i'+1)i'}$, $q_{ij,(i+1)(j+1)} = \frac{p_0}{\Delta}q_{i'i',(i'+1)(i'+1)}$ for all $0 \leq i' < P$. Thus, $q_{11} = \frac{\Delta}{p_0}q_{10,21}$. Moreover, by making use of the property
\[
\frac{p_r(1-p_0)}{p_0}q_{ij,(i+1)j} = q_{i'i',(i'+1)i'} - q_{ij,(i+1)j},
\frac{p_r(1-p_0)}{p_0}q_{ij,(i+1)(j+1)} = q_{i'i',(i'+1)(i'+1)} - q_{ij,(i+1)(j+1)}
\]
for $0 \leq j < i < P$ and $0 \leq i' < P$, one may readily verify that for $1 < i < P$,
\begin{equation}
\label{eqn:appendix_qii_expression_1}
q_{ii} = \frac{\Delta}{p_0}q_{10,(i+1)i} + \frac{p_r(1-p_0)}{p_0}\sum\nolimits_{j = 1}^{i-1} q_{jj}q_{10,(i-j+1)(i-j)}
\end{equation}

On the other hand, for $0 < i < P-1$, $q_{(i+1)(i+1)}$ can also be expressed as
\begin{align}
\label{eqn:appendix_qii_expression_2}
q_{(i+1)(i+1)} = &\frac{p_r}{\Delta}q_{ii} + \sum\nolimits_{j = 0}^{i-1} \frac{p_r(1-p_0)(1-p_r)}{\Delta}q_{jj}q_{(j+1)j, (i+1)i} \nonumber \\
= &\frac{p_r}{\Delta}q_{ii} + \sum\nolimits_{j = 0}^{i-1} \frac{p_r(1-p_0)(1-p_r)}{\Delta}q_{jj}q_{10,(i-j+1)(i-j)}
\end{align}
where the last equality holds due to $q_{ij,(i+i')(j+j')} = q_{10,(i'+1)j'}$ for all $i,i',j,j'$ subject to $0 \leq j < i$ and $j+j' < i+i' < P$. Then, the addition of \eqref{eqn:appendix_qii_expression_2} and \eqref{eqn:appendix_qii_expression_1} multiplied by $-\frac{p_0(1-p_r)}{\Delta}$ on both sides yields $q_{(i+1)(i+1)} - q_{ii} = - \frac{p_0(1-p_r)}{\Delta}q_{10,(i+1)i}$, which implies for $1 < i < P$,
\begin{equation}
\label{eqn:q_ii_difference_formula}
\frac{1-p_0}{p_0}(q_{ii} - q_{(i-1)(i-1)}) = (1 - \frac{1}{\Delta})q_{10,i(i-1)}.
\end{equation}

We last characterize $q_{10, i(i-1)}$ given that $1 < i < P$. Notice that $q_{ij,(i+1)(j+1)} = \frac{p_0p_r(1-p_0)(1-p_r)}{\Delta^2} + \frac{p_0p_r}{\Delta} = \frac{p_0p_r}{\Delta^2}$ for $0 \leq j < i$. It turns out that
\begin{equation}
\label{eqn:appendix_q_10_ii-1_expression}
q_{10,i(i-1)} = \frac{(p_0p_r)^{i-1}}{\Delta^{2(i-1)}}\sum\nolimits_{j = 1}^{i-1}N_{i-1,j}[(1-p_0)(1-p_r)]^{i-j-1} = \frac{(p_0p_r)^{i-1}}{\Delta^{2(i-1)}}\sum\nolimits_{j = 1}^{i-1}N_{i-1,j}(1-\Delta)^{i-j-1}
\end{equation}
where $N_{i-1,j}$ represents the number of all those transitions from $(1,0)$ to $(i, i-1)$ that contain
\begin{itemize}
\item exactly $j$ $1$-step transitions in the form $(i',j') \rightarrow (i'+1,j'+1)$, $i-j-1$ $1$-step transitions in the form $(i',j') \rightarrow (i'+1,j')$, and $i-j-1$ $1$-step transitions in the form $(i', j') \rightarrow (i', j'+1)$;
\item no $2$-step transitions in the form $(i',j') \rightarrow (i'+1,j') \rightarrow (i'+1,j'+1)$.
\end{itemize}
Under this constraint, $N_{i-1,j}$ coincides with the Narayana number (see, e.g., Chapter 2 in \cite{Petersen}) with parameters $1 \leq j \leq i-1$, so $N_{i-1,j} = \frac{1}{i-1}\binom{i-1}{j}\binom{i-1}{j-1}$. As $N_{i-1,j} = N_{i-1,i-j}$,
\begin{align}
\label{eqn:appendix_combinatorial_analysis1}
&\sum\nolimits_{j = 1}^{i-1}N_{i-1,j}(1 - \Delta)^{i-j-1} \nonumber \\
= &\sum\nolimits_{j = 1}^{i-1}N_{i-1,j}(1 - \Delta)^{j-1} \nonumber \\
= &\sum\nolimits_{j = 1}^{i-1}N_{i-1,j}\sum\nolimits_{j' = 1}^{j} (-\Delta)^{j'-1}\binom{j-1}{j'-1} \nonumber \\
= &\sum\nolimits_{j'=1}^{i-1} (-\Delta)^{j'-1} \sum\nolimits_{j = j'}^{i-1} \binom{j-1}{j'-1}N_{i-1,j} \nonumber \\
= &\sum\nolimits_{j'=1}^{i-1} \frac{(-\Delta)^{j'-1}}{i-1} \binom{i-1}{j'-1}\binom{2i-j'-1}{i} \nonumber \\
= &\sum\nolimits_{j=1}^{i-1} \frac{(-\Delta)^{i-j-1}}{i-1} \binom{i-1}{j}\binom{i+j-1}{i},
\end{align}
where the second last equality can be readily verified based on the combinatorial equation $\binom{n+m}{n+1} = \sum_{j=n-m+1}^n \binom{n}{j}\binom{m}{n+1-j}$ for $1 \leq m \leq n+1$. By plugging \eqref{eqn:appendix_combinatorial_analysis1} back to \eqref{eqn:appendix_q_10_ii-1_expression},
\[
q_{10,i(i-1)} = \left(\frac{p_0p_r}{\Delta^2}\right)^{i-1}\sum_{j=1}^{i-1}\frac{(-\Delta)^{i-j-1}}{i-1} \binom{i-1}{j}\binom{i+j-1}{i}.
\]
Thus, the right-hand side of \eqref{eqn:q_ii_difference_formula} can be expressed as
\begin{align}
\label{eqn:appendix_q_10_i_i-1_formula1}
&(1 - \frac{1}{\Delta})q_{10,i(i-1)}  \nonumber\\
= & \sum\nolimits_{j=1}^{i-1} \frac{(p_0p_r)^{i-1}}{(i-1)(-\Delta)^{i+j-1}} \binom{i-1}{j}\binom{i+j-1}{i} + \sum\nolimits_{j=1}^{i-1} \frac{(p_0p_r)^{i-1}}{(i-1)(-\Delta)^{i+j}} \binom{i-1}{j}\binom{i+j-1}{i}
\end{align}
As $\binom{i-1}{j}= \frac{i-1}{j}\binom{i-2}{j-1}$, the first term in \eqref{eqn:appendix_q_10_i_i-1_formula1} can be expressed as
\begin{align*}
&\sum\nolimits_{j=1}^{i-1} \frac{(p_0p_r)^{i-1}}{(i-1)(-\Delta)^{i+j-1}} \binom{i-1}{j}\binom{i+j-1}{i} \\
= &\sum\nolimits_{j=1}^{i-1}\frac{(p_0p_r)^{i-1}}{j(-\Delta)^{i+j-1}} \binom{i-2}{j-1}\binom{i+j-1}{i} \\
=& \frac{(p_0p_r)^{i-1}}{(-\Delta)^i} + \sum\nolimits_{j=2}^{i-1} \frac{(p_0p_r)^{i-1}}{j(-\Delta)^{i+j-1}} \binom{i-2}{j-1}\binom{i+j-1}{i}
\end{align*}
Moreover, by altering the index of $j$ and the fact $\frac{1}{i-1}\binom{2i-2}{i} = \frac{1}{i}\binom{2i-2}{i-1}$, we can express the second term in \eqref{eqn:appendix_q_10_i_i-1_formula1} as
\begin{align*}
& \sum\nolimits_{j=1}^{i-1} \frac{(p_0p_r)^{i-1}}{(i-1)(-\Delta)^{i+j}} \binom{i-1}{j}\binom{i+j-1}{i} \\
= &\sum\nolimits_{j=2}^{i-1} \frac{(p_0p_r)^{i-1}}{(i-1)(-\Delta)^{i+j-1}} \binom{i-1}{j-1}\binom{i+j-2}{i} + \frac{(p_0p_r)^{i-1}}{i(-\Delta)^{2i-1}}\binom{2i-2}{i-1}.
\end{align*}
Consequently,
\begin{align}
\label{eqn:appendix_q_10_i_i-1_formula3}
&(1 - \frac{1}{\Delta})q_{10,i(i-1)} \nonumber \\
= & \frac{(p_0p_r)^{i-1}}{(-\Delta)^i} + \sum\nolimits_{j=2}^{i-1} \frac{(p_0p_r)^{i-1}}{j(-\Delta)^{i+j-1}} \binom{i-2}{j-1}\binom{i+j-1}{i} \nonumber \\
& + \sum\nolimits_{j=2}^{i-1} \frac{(p_0p_r)^{i-1}}{(i-1)(-\Delta)^{i+j-1}} \binom{i-1}{j-1}\binom{i+j-2}{i} + \frac{(p_0p_r)^{i-1}}{i(-\Delta)^{2i-1}}\binom{2i-2}{i-1} \nonumber\\
=&\frac{(p_0p_r)^{i-1}}{(-\Delta)^i} + \sum\nolimits_{j=2}^{i-1} \frac{(p_0p_r)^{i-1}}{j(-\Delta)^{i+j-1}} \binom{i+j-2}{i-1}\binom{i-1}{j-1} + \frac{(p_0p_r)^{i-1}}{i(-\Delta)^{2i-1}}\binom{2i-2}{i-1}  \nonumber \\%
= &\sum\nolimits_{j=1}^{i} \frac{(p_0p_r)^{i-1}}{j(-\Delta)^{i+j-1}} \binom{i+j-2}{i-1}\binom{i-1}{j-1}  \nonumber \\
= &\sum\nolimits_{j=0}^{i-1} \frac{(p_0p_r)^{i-1}T_{i-1,j}}{(-\Delta)^{i+j}}.
\end{align}
where the second equality holds because $\frac{1}{j}\binom{i-2}{j-1}\binom{i+j-1}{i} + \frac{1}{i-1}\binom{i-1}{j-1}\binom{i+j-2}{i} = \frac{1}{j}\binom{i+j-2}{i-1}\binom{i-1}{j-1}$.

Eq. \eqref{eqn:appendix_equivalent_condition} can now be verified based on \eqref{eqn:q_ii_difference_formula} and \eqref{eqn:appendix_q_10_i_i-1_formula3} with the setting $i = P - 1$. %

\subsection{ Proof of Theorem \ref{thm:E[D_P]_recursive_formula} }
\label{Appendix:proof_of Theorem6}
Given that there are $R = 2$ receivers and $P+1$ source packets in the network. %
For the parameter $E_{\max}$ defined in \eqref{eqn:E_max_definition}, by the min-max identity, $E_{\max} = 1/p_1 + 1/p_2 - E_{\min}$, where $E_{\min} = \mathbb{E}[\min\{N_1, N_2\}] = 1/(1-(1-p_1)(1-p_2)) = 1/(p_1 + p_2 - p_1p_2)$. %
Moreover, we have
\[
\begin{split}
&\mathrm{Pr}(T_{P+1,2} \leq T_{P,1})+\mathrm{Pr}(T_{P+1,1} > T_{P,2},T_{P+1,2} > T_{P,1}) = \mathrm{Pr}(T_{P+1,1} > T_{P,2}), \\
&\mathrm{Pr}(T_{P+1,1} \leq T_{P,2})+\mathrm{Pr}(T_{P+1,1} > T_{P,2},T_{P+1,2} > T_{P,1}) = \mathrm{Pr}(T_{P+1,2} > T_{P,1}),
\end{split}
\]
and taking a summation on each side of the above two equations yields
\begin{equation}
\begin{split}
&1 + \mathrm{Pr}(T_{P+1,1} > T_{P,2},T_{P+1,2} > T_{P,1}) = \mathrm{Pr}(T_{P+1,1} > T_{P,2}) + \mathrm{Pr}(T_{P+1,2} > T_{P,1}).
\end{split}
\end{equation} %
As a result, Eq. \eqref{eqn:recursive_expression_of_fullduplex1} in Lemma \ref{lemma:recursive_expression_of_fullduplex1} and \eqref{eqn:recursive_eq_of_wb} can be respectively rewritten as
\begin{align*}
\mathbb{E}[D_{P+1}] - \mathbb{E}[D_{P}]
 =& \mathrm{Pr}(T_{P+1,1} > T_{P,2})(\frac{1}{p_1}-E_{\min}) + \mathrm{Pr}(T_{P+1,2} > T_{P,1})(\frac{1}{p_2}-E_{\min}) + E_{\min} + \\
 & \mathrm{Pr}(S_{P+1} > \max\{T_{P,1}, T_{P,2}\})(\frac{1}{p_0}-1),
\end{align*}
\[
\mathbb{E}[\hat{D}_{P+1}] - \mathbb{E}[\hat{D}_{P}]
 = \mathrm{Pr}(\hat{T}_{P+1,1} > \hat{T}_{P,2})(\frac{1}{p_1}-E_{\min}) + \mathrm{Pr}(\hat{T}_{P+1,2} > \hat{T}_{P,1})(\frac{1}{p_2}-E_{\min}) + E_{\min}.
\]
By the above two equations and \eqref{eqn:epsilon_identity}, in order to prove \eqref{eqn:equivalent_recursive_expression_of_E[D_P]}, it suffices to show
\begin{equation}
\label{eqn:varepsilon_equivalent_expression}
\begin{split}
\varepsilon_{P+1}
=&(\mathrm{Pr}(T_{P+1,1} > T_{P,2}) - \mathrm{Pr}(\hat{T}_{P+1,1} > \hat{T}_{P,2}))(\frac{1}{p_1}-E_{\min}) +\\
&(\mathrm{Pr}(T_{P+1,2} > T_{P,1}) - \mathrm{Pr}(\hat{T}_{P+1,2} > \hat{T}_{P,1}))(\frac{1}{p_2}-E_{\min})
\end{split}
\end{equation}
Observe that
\begin{align}
\label{eqn:T_P+1_T_P_equiv_comparison}
\mathrm{Pr}(T_{P+1,1} > T_{P,2})
= &1 - \mathrm{Pr}(T_{P,2} \geq T_{P+1,1}) \nonumber \\
= &1 - \mathrm{Pr}(T_{P+1,2} > T_{P+1,1}) + \mathrm{Pr}(T_{P+1,2} > T_{P+1,1} > T_{P,2}).
\end{align}
Due to the memoryless property of geometric distribution, under the condition $T_{P+1,2} > T_{P,1}$ and $T_{P+1,1} > T_{P, 2}$, the events $T_{P+1,2} > T_{P+1,1}$ and $T_{P+1,2} = T_{P+1,1}$ are independent of the arrival timeslot $S_{P+1}$ of packet $P+1$ at the RS, and they have respective probability $\frac{p_1(1-p_2)}{p_1+p_2-p_1p_2}$ and $\frac{p_1p_2}{p_1+p_2-p_1p_2}$ to occur. Hence,
\begin{equation*}
\begin{split}
&\mathrm{Pr}(T_{P+1,2} > T_{P+1,1} > T_{P,2})\\
=&\mathrm{Pr}(T_{P+1,2} > T_{P+1,1} | T_{P+1,2} > T_{P,1}, T_{P+1,1} > T_{P, 2}){Pr}(T_{P+1,2} > T_{P,1}, T_{P+1,1} > T_{P, 2}) \\
= &\mathrm{Pr}(T_{P+1,2} > T_{P,1}, T_{P+1,1} > T_{P, 2})\frac{p_1(1-p_2)}{p_1+p_2-p_1p_2}, \\
&\mathrm{Pr}(T_{P+1,2} = T_{P+1,1})= \mathrm{Pr}(T_{P+1,2} > T_{P,1}, T_{P+1,1} > T_{P, 2})\frac{p_1p_2}{p_1+p_2-p_1p_2}.
\end{split}
\end{equation*}
It turns out that $\mathrm{Pr}(T_{P+1,2} > T_{P+1,1} > T_{P,2}) = \mathrm{Pr}(T_{P+1,2} = T_{P+1,1})\frac{1-p_2}{p_2}$. %
By plugging the above expression back to \eqref{eqn:T_P+1_T_P_equiv_comparison}, we obtain
\begin{equation}
\label{eqn:Appendix-B-last-1}
\mathrm{Pr}(T_{P+1,1} > T_{P,2}) = 1 - \mathrm{Pr}(T_{P+1,2} > T_{P+1,1}) + \mathrm{Pr}(T_{P+1,2} = T_{P+1,1})\frac{1-p_2}{p_2},
\end{equation}
and similarly
\begin{equation}
\label{eqn:Appendix-B-last-2}
\mathrm{Pr}(T_{P+1,2} > T_{P,1}) = 1 - \mathrm{Pr}(T_{P+1,1} > T_{P+1,2}) + \mathrm{Pr}(T_{P+1,1} = T_{P+1,2})\frac{1-p_1}{p_1}.
\end{equation}
Based on \eqref{eqn:Appendix-B-last-1}, \eqref{eqn:Appendix-B-last-2}, and $E_{\min} = 1/(p_1 + p_2 - p_1p_2)$, the correctness of \eqref{eqn:varepsilon_equivalent_expression} can be verified.

\subsection{ Proof of Theorem \ref{thm:E_DP_approximation} }
\label{Proof of Theorem_E_DP_approximation}
It remains to prove $\sum\nolimits_{j = 2}^P \varepsilon_{j} \geq 0$. To ease the analysis, we make use of the following Markov chain $\mathcal{M}$ consisting of $(P+1)^2$ states, in which (i) state $(i,j)$, $0 \leq i, j \leq P$, represents the scenario that receivers $1$ and $2$ have respectively obtained $i$ and $j$ packets; (ii) there is a $1$-step transition once at least one receiver obtains a new packet. In the Markov chain, let $p_{ij,i'j'}$ represent the $1$-step transition probability from state $(i,j)$ to $(i',j')$, and $q_{ij}$ denote the total probability to visit state $(i,j)$ among all paths from $(0,0)$ to $(P, P)$. %
Notice that the Markov chain $\mathcal{M}$ defined herein is different from the one modeled at the beginning of Sec. III-C and illustrated in Fig. \ref{fig:markov_chain}, because it does not involve the number of received packets at the RS in the state description. Observe that for $1 \leq i < P$,
\begin{equation}
\label{eqn:q_ii_interpretation}
q_{ii} = \mathrm{Pr}(T_{i+1,1} > T_{i,2}, T_{i+1,2} > T_{i,1}).
\end{equation}
For brevity, write
$Q_{0}^i = q_{(i-1)(i-1)}p_{(i-1)(i-1), ii},~Q_{1}^{i} = \sum\nolimits_{j=0}^{i-1} q_{ij},~Q_2^i = \sum\nolimits_{j=0}^{i-1} q_{ji}$ for $1 \leq i \leq P$ %
and write $\Delta_{ij} = 1 - (1-p_i)(1-p_j) = p_i + p_j - p_ip_j$ for $0 \leq i < j \leq 2$. In this way,
\begin{equation}
Q_{0}^i = \mathrm{Pr}(T_{i,1} = T_{i,2}),~
Q_1^i = \mathrm{Pr}(T_{i,2} > T_{i,1}),~
Q_2^i = \mathrm{Pr}(T_{i,1} > T_{i,2}) ,~
Q_0^i + Q_1^i + Q_2^i = 1,
\end{equation}
\begin{equation}
\label{eqn:varepsilon_alternative_expression}
\varepsilon_{i}
=(Q_0^i - \hat{Q}_0^i)\frac{(1-p_1)(1-p_2)(p_1+p_2)}{p_1p_2\Delta_{12}} + (\hat{Q}_2^i - Q_2^i)\frac{p_1(1-p_2)}{p_2\Delta_{12}} + (\hat{Q}_1^i - Q_1^i)\frac{(1-p_1)p_2}{p_1\Delta_{12}}
\end{equation}
for $1 \leq i \leq P$. We next characterize the $1$-step transition probability in $\mathcal{M}$.

For $0 \leq i < P$, by the memoryless property of a geometric distribution, the $1$-step transition probability starting from state $(i,i)$ is not affected by the arrival time $S_{i+1}$ of the $(i+1)^{st}$ packet at the RS and thus is invariant of $i$. Specifically,
\begin{equation}
\label{eqn:step_transition_prob_state_ii}
p_{ii,(i+1)(i+1)} = \frac{p_1p_2}{\Delta_{12}},\quad
p_{ii,(i+1)i} = \frac{p_1(1-p_2)}{\Delta_{12}},\quad
p_{ii,i(i+1)} = \frac{(1-p_1)p_2}{\Delta_{12}}.
\end{equation}

For $0 \leq j < i < P$, the $1$-step transition probability starting from state $(i,j)$ is affected by the arrival time $S_{i+1}$ of the $(i+1)^{st}$ packet at the RS. Notice that when $\mathcal{M}$ is in state $(i, j)$, it implies the occurrence of the joint event of $T_{i+1,1} > T_{j,2}$ and $T_{j+1, 2} > T_{i,1}$, which will be denoted by $A_{ij}$. Thus, we can respectively express $p_{ij,i(j+1)} = \mathrm{Pr}(T_{j+1, 2} < T_{i+1, 1} | A_{ij})$, $p_{ij,(i+1)(j+1)} = \mathrm{Pr}(T_{j+1, 2} = T_{i+1, 1} | A_{ij})$, and $p_{ij,(i+1)j}
= \mathrm{Pr}(T_{j+1, 2} > T_{i+1, 1} | A_{ij})$ as
\begin{align}
\label{eqn:transition_prob_2D_Markov_in_proof_1}
p_{ij,i(j+1)}
= &\mathrm{Pr}(T_{j+1, 2} < S_{i+1} | S_{i+1} > T_{i,1}, A_{ij})\mathrm{Pr}(S_{i+1} > T_{i,1} | A_{ij})+ \nonumber\\
  &\mathrm{Pr}(S_{i+1} \leq T_{j+1, 2} < T_{i+1, 1} | S_{i+1} > T_{i,1}, A_{ij}) \nonumber \mathrm{Pr}(S_{i+1} > T_{i,1} | A_{ij})+\nonumber\\
  &\mathrm{Pr}(T_{j+1, 2} < T_{i+1, 1} | S_{i+1} \leq T_{i,1}, A_{ij})\mathrm{Pr}(S_{i+1} \leq T_{i,1} | A_{ij})\nonumber \\
= & \frac{(1-p_1)p_2}{\Delta_{12}} + \frac{(1-p_0)p_1p_2}{\Delta_{02}\Delta_{12}}\mathrm{Pr}(S_{i+1} > T_{i,1} | A_{ij}) \nonumber\\
= &\frac{(1-p_1)p_2}{\Delta_{12}}(1 + \frac{p_1}{1-p_1}\alpha_{ij}), \\
\label{eqn:transition_prob_2D_Markov_in_proof_2}
p_{ij,(i+1)(j+1)}
= &\mathrm{Pr}(S_{i+1} \leq T_{j+1, 2} = T_{i+1, 1} | S_{i+1} > T_{i,1}, A_{ij}) \mathrm{Pr}(S_{i+1} > T_{i,1} | A_{ij})+\nonumber\\
  &\mathrm{Pr}(T_{j+1, 2} = T_{i+1, 1} | S_{i+1} \leq T_{i,1}, A_{ij})\mathrm{Pr}(S_{i+1} \leq T_{i,1} | A_{ij}) \nonumber\\
= & \frac{p_1p_2}{\Delta_{12}} -  \frac{(1-p_0)p_1p_2^2}{\Delta_{02}\Delta_{12}}\mathrm{Pr}(S_{i+1} > T_{i,1} | A_{ij}) \nonumber\\
= &\frac{p_1p_2}{\Delta_{12}}(1 - p_2\alpha_{ij}), \\
\label{eqn:transition_prob_2D_Markov_in_proof_3}
p_{ij,(i+1)j}
= &\mathrm{Pr}(S_{i+1} \leq T_{i+1, 1} < T_{j+1, 2} | S_{i+1} > T_{i,1}, A_{ij})\mathrm{Pr}(S_{i+1} > T_{i,1} | A_{ij})+\nonumber\\
  &\mathrm{Pr}(T_{j+1, 2} > T_{i+1, 1} | S_{i+1} \leq T_{i,1}, A_{ij})\mathrm{Pr}(S_{i+1} \leq T_{i,1} | A_{ij})\nonumber \\
= & \frac{p_1(1-p_2)}{\Delta_{12}} - \frac{(1-p_0)p_1p_2(1-p_2)}{\Delta_{02}\Delta_{12}}\mathrm{Pr}(S_{i+1} > T_{i,1} | A_{ij}) \nonumber\\
= &\frac{p_1(1-p_2)}{\Delta_{12}}(1 - p_2\alpha_{ij}),
\end{align}
where $\alpha_{ij} = \left\{ \begin{matrix}
\frac{1-p_0}{\Delta_{02}}\mathrm{Pr}(S_{i+1} > T_{i,1} | A_{ij}) & \mathrm{if}~i>j \\
\frac{1-p_0}{\Delta_{01}}\mathrm{Pr}(S_{j+1} > T_{j,2} | A_{ij}) & \mathrm{if}~i<j \\
\end{matrix} \right.$.

Similarly, for $0 \leq i < j \leq P$,
\begin{equation}
\begin{split}
\label{eqn:transition_prob_2D_Markov_in_proof_4}
p_{ij,(i+1)j} &= \frac{p_1(1-p_2)}{\Delta_{12}}(1 + \frac{p_2}{1-p_2}\alpha_{ij}),
p_{ij,(i+1)(j+1)} = \frac{p_1p_2}{\Delta_{12}}(1 - p_1\alpha_{ij}), \\
p_{ij,i(j+1)} &= \frac{(1-p_1)p_2}{\Delta_{12}}(1 - p_1\alpha_{ij}).
\end{split}
\end{equation}

For the special case $p_0 = 1$, let $\hat{p}_{ij,i'j'}$, $\hat{q}_{ij}$ and $\hat{Q}^i_j$ respectively represent $p_{ij,i'j'}$, $q_{ij}$ and $Q^i_j$. Based on the above derivation of $1$-step transition probabilities in $\mathcal{M}$, we obtain the following comparisons. For $0 \leq i < P$,
\begin{equation}
\label{eqn:p_ij_hatp_ij_comparison_0}
p_{ii,(i+1)i} = \hat{p}_{ii,(i+1)i},\quad
p_{ii,i(i+1)} = \hat{p}_{ii,i(i+1)},\quad
p_{ii,(i+1)(i+1)} = \hat{p}_{ii,(i+1)(i+1)}.
\end{equation}
For $0 \leq j < i < P$,
\begin{equation}
\label{eqn:p_ij_hatp_ij_comparison_1}
p_{ij,i(j+1)} - \hat{p}_{ij,i(j+1)} \geq 0,~
p_{ij,(i+1)(j+1)} - \hat{p}_{ij,(i+1)(j+1)} \leq 0,~
p_{ij,(i+1)j} - \hat{p}_{ij,(i+1)j} \leq 0.
\end{equation}
For $0 \leq i < j < P$,
\begin{equation}
\label{eqn:p_ij_hatp_ij_comparison_2}
p_{ij,(i+1)j} - \hat{p}_{ij,(i+1)j} \geq 0,~
p_{ij,(i+1)(j+1)} - \hat{p}_{ij,(i+1)(j+1)} \leq 0,~
p_{ij,i(j+1)} - \hat{p}_{ij,i(j+1)} \leq 0.
\end{equation}

Assume $2 \leq i \leq P$ and $p_0 < 1$. Eqs. \eqref{eqn:p_ij_hatp_ij_comparison_0}-\eqref{eqn:p_ij_hatp_ij_comparison_2} together imply that there are higher probabilities to visit state $(i,i)$ along the paths from $(0,0)$ to $(P,P)$ compared with the case $p_0 = 1$, that is, $q_{(i-1)(i-1)} > \hat{q}_{(i-1)(i-1)}$.
Since $p_{(i-1)(i-1),ii} = \hat{p}_{(i-1)(i-1),ii}$, $Q_{0}^i > \hat{Q}_{0}^i$.
Since $Q_0^i + Q_1^i + Q_2^i = 1$,
\begin{equation}
\label{eqn:Q1+Q2_inequality}
\hat{Q}_1^i + \hat{Q}_2^i > Q_1^i + Q_2^i.
\end{equation}

When $p_1 = p_2$, we have $Q_1^i = Q_2^i$, so \eqref{eqn:Q1+Q2_inequality} implies $\hat{Q}_1^i > Q_1^i$ and $\hat{Q}_2^i > Q_2^i$. Therefore, $\varepsilon_i > 0$ and $\sum_{2 \leq i \leq P}\varepsilon_i > 0$. %

Assume $p_1 < p_2$. In this case, $Q_0^i > \hat{Q}_0^i$, $\hat{Q}_2^i > Q_2^i > Q_1^i$, and $\hat{Q}_2^i$ converges to $1$ while $\hat{Q}_0^i$ and $\hat{Q}_1^i$ converge to $0$ with increasing $i$ (assume $P$ is sufficiently large). For relatively small $i$, $\hat{Q}_1^i$ is still larger than $Q_1^i$ due to the effect of \eqref{eqn:p_ij_hatp_ij_comparison_1}, so we have $\varepsilon_i > 0$. However, with increasing $i$, $\hat{Q}_1^i < Q_1^i$ will occur because $\hat{Q}_1^i$ decreases faster than $Q_1^i$, ($Q_1^i$ may or may not converge to $0$ depending on whether $p_0$ is larger than $p_1$). %
As a result, for large $P$, \eqref{eqn:Q1+Q2_inequality} is insufficient to imply $\varepsilon_i > 0$ so that we need further manipulation on the expression of $\varepsilon_i$. %

For $1 \leq i < P$, write $Q_{10}^i = q_{i(i-1)}p_{i(i-1),ii}, Q_{20}^i = q_{(i-1)i}p_{(i-1)i,ii}$, %
and let $\hat{Q}_{10}^i$, $\hat{Q}_{20}^i$ respectively represent $Q_{10}^i$, $Q_{20}^i$ for the special case $p_0 = 1$. %
In terms of $Q_{10}^i$, we can express $Q_1^{i+1}$, $1 \leq i < P$, recursively as
\begin{equation}
\label{eqn:Q1_Q10_recursive_formula}
Q_1^{i+1}
= Q_1^{i} - Q_{10}^{i} + q_{ii}p_{ii,(i+1)i}
= Q_{1}^{i} - Q_{10}^{i} + q_{ii}\frac{p_1(1-p_2)}{\Delta_{12}}
= Q_1^{i} - Q_{10}^{i} + Q_0^{i+1}\frac{1-p_2}{p_2},
\end{equation}
where the last equality holds by \eqref{eqn:step_transition_prob_state_ii} and the definition of $Q_0^{i}$. %
Similarly,
\begin{equation}
\label{eqn:Q2_Q20_recursive_formula}
Q_2^{i+1} = Q_2^{i} - Q_{20}^{i} + q_{ii}\frac{(1-p_1)p_2}{\Delta_{12}}
= Q_2^{i} - Q_{20}^{i} + Q_0^{i+1}\frac{1-p_1}{p_1}.
\end{equation}
By plugging \eqref{eqn:Q1_Q10_recursive_formula}, \eqref{eqn:Q2_Q20_recursive_formula} back to \eqref{eqn:varepsilon_alternative_expression}, we can deduce the following recursive expression of $\varepsilon_{i+1}$,
\begin{equation*}
\begin{split}
& \varepsilon_{i+1} \\
=& (\hat{Q}_2^{i}-Q_2^{i}+Q_{20}^{i}-\hat{Q}_{20}^{i})\frac{p_1(1-p_2)}{p_2\Delta_{12}} + (\hat{Q}_1^{i}-Q_1^{i}+Q_{10}^{i}-\hat{Q}_{10}^{i})\frac{(1-p_1)p_2}{p_1\Delta_{12}} \\
=& \varepsilon_{i} + (Q_{20}^{i}-\hat{Q}_{20}^{i})\frac{p_1(1-p_2)}{p_2\Delta_{12}} + (Q_{10}^{i}-\hat{Q}_{10}^{i})\frac{(1-p_1)p_2}{p_1\Delta_{12}} -(Q_0^{i} - \hat{Q}_0^{i})\frac{(1-p_1)(1-p_2)(p_1+p_2)}{p_1p_2\Delta_{12}}, \\
\end{split}
\end{equation*}
where $C_{1}^{i} = Q_{10}^{i}-Q_0^{i}\frac{1-p_2}{p_2} = Q_{10}^{i} - q_{(i-1)(i-1)}\frac{p_1(1-p_2)}{\Delta_{12}}$,
$C_{2}^{i} = Q_{20}^{i}-Q_0^{i}\frac{1-p_1}{p_1} = Q_{20}^{i} - q_{(i-1)(i-1)}\frac{(1-p_1)p_2}{\Delta_{12}}$. %
In this way, for $1 \leq i < P$, we have $\varepsilon_{i+1} = \varepsilon_{i} + (C_{1}^{i} - \hat{C}_1^{i})\frac{(1-p_1)p_2}{p_1\Delta_{12}} + (C_{2}^{i} - \hat{C}_2^{i})\frac{p_1(1-p_2)}{p_2\Delta_{12}}$,
where $\hat{C}_1^{i}$, $\hat{C}_2^{i}$ respectively refer to $C_{1}^{i}$, $C_{2}^{i}$ in the case of $p_0 = 1$. Because
\begin{align}
\label{eqn:C1+C2_q_ii}
C_1^i + C_2^i = &Q_{10}^i + Q_{20}^i - q_{(i-1)(i-1)}\frac{p_1+p_2 - 2p_1p_2}{\Delta_{12}} = q_{ii} - q_{(i-1)(i-1)},
\end{align}
we has $\varepsilon_{i+1} =  (q_{ii} - \hat{q}_{ii})\frac{p_1(1-p_2)}{p_2\Delta_{12}} + \sum_{j=1}^{i} (C_1^j - \hat{C}_1^j)(\frac{1}{p_1} - \frac{1}{p_2})$. %
When $p_0 = 1$ or $p_1 = p_2 = 1$, $q_{ii} = \hat{q}_{ii}$ and $C_1^j = \hat{C}_1^j$, so that $\varepsilon_{i+1} = 0$ for all $1 \leq i < P$.%

Assume $p_0, p_1, p_2 \neq 1$. Based on \eqref{eqn:transition_prob_2D_Markov_in_proof_1} and \eqref{eqn:transition_prob_2D_Markov_in_proof_4}, we have
\begin{align*}
q_{11} - \hat{q}_{11}
= \frac{p_1p_2(1-p_1)(1-p_2)}{\Delta_{12}^2}(\frac{p_1}{1-p_1}\alpha_{01}+\frac{p_2}{1-p_2}\alpha_{10}) > 0, \\
C_{1}^1 - \hat{C}_1^1
= Q_{10}^1 - \hat{Q}_{10}^1 = \frac{p_1p_2(1-p_1)(1-p_2)}{\Delta_{12}^2}\frac{p_2}{1-p_2}\alpha_{10} > 0,
\end{align*}
so that $\varepsilon_{2} > 0$. %
With increasing $i$, both $q_{ii}$ and $\hat{q}_{ii}$ decrease and $q_{ii} - \hat{q}_{ii}$ converges to a nonnegative constant value (for sufficiently large $P$). We can then deduce, based on \eqref{eqn:C1+C2_q_ii}, that both $C_1^j$ and $C_2^j$ converge to zero with increasing $i$ too. As a result, even if $C_1^i - \hat{C}_1^i < 0$ is possible to occur, $|C_1^i - \hat{C}_1^i|$ is negligible compared with $\sum_{j=1}^i (q_{jj} - \hat{q}_{jj}) > 0$. We can now assert $\sum\nolimits_{j = 2}^P \varepsilon_{j} > 0$.
\section*{Acknowledgement}
The authors would like to appreciate the valuable suggestions by the editor as well as anonymous reviewers
to help improve the quality of the paper.
\linespread{1.2}

\end{document}